\documentclass[twocolumn]{aastex62}

\usepackage{amsmath}

\usepackage{xcolor, fontawesome, microtype}
\definecolor{twitterblue}{RGB}{64,153,255}
\definecolor{linkcolor}{rgb}{0.1216,0.4667,0.7059}

\usepackage{ mathrsfs }
\usepackage[T1]{fontenc}

\graphicspath{{./figures/}{}}

\received{April 29, 2020}
\revised{June 26, 2020}
\accepted{June 28, 2020}
\submitjournal{AJ}

\shorttitle{Planet Occurrence in the Campaign 5 FGK Sample}

\shortauthors{Zink et al. (2020)}

\begin{document}

\title{Scaling \emph{K2}. III. Comparable Planet Occurrence in the FGK Samples of Campaign 5 and Kepler}

\correspondingauthor{Jon Zink}
\email{jzink@astro.ucla.edu}

\author[0000-0003-1848-2063]{Jon K. Zink}
\affiliation{Department of Physics and Astronomy, University of California, Los Angeles, CA 90095}

\author[0000-0003-3702-0382]{Kevin K. Hardegree-Ullman}
\affiliation{Caltech/IPAC-NASA Exoplanet Science Institute, Pasadena, CA 91125}

\author[0000-0002-8035-4778]{Jessie L. Christiansen}
\affiliation{Caltech/IPAC-NASA Exoplanet Science Institute, Pasadena, CA 91125}

\author[0000-0003-0967-2893]{Erik A. Petigura}
\affiliation{Department of Physics and Astronomy, University of California, Los Angeles, CA 90095}

\author[0000-0001-8189-0233]{Courtney D. Dressing}
\affiliation{Department of Astronomy, University of California, Berkeley, CA 94720}

\author[0000-0001-5347-7062]{Joshua E. Schlieder}
\affiliation{Exoplanets and Stellar Astrophysics Laboratory, Code 667, NASA Goddard Space Flight Center, Greenbelt, MD 20771}

\author[0000-0002-5741-3047]{David R. Ciardi}
\affiliation{Caltech/IPAC-NASA Exoplanet Science Institute, Pasadena, CA 91125}

\author[0000-0002-1835-1891]{Ian J. M. Crossfield}
\affiliation{Department of Physics, and Kavli Institute for Astrophysics and Space Research, Massachusetts Institute of Technology, Cambridge, MA 02139}
\affiliation{Department of Physics and Astronomy, University of Kansas, Lawrence, KS 66045 }

\begin{abstract}
Using our \emph{K2} Campaign 5 fully automated planet detection data set (43 planets), which has corresponding measures of completeness and reliability, we infer an underlying planet population model for the FGK dwarfs sample (9,257 stars). Implementing a broken power-law for both the period and radius distribution, we find an overall planet occurrence of $1.00^{+1.07}_{-0.51}$ planets per star within a period range of 0.5--38 days. Making similar cuts and running a comparable analysis on the \emph{Kepler} sample (2,318 planets; 94,222 stars), we find an overall occurrence of $1.10\pm0.05$ planets per star. Since the Campaign 5 field is nearly 120 angular degrees away from the \emph{Kepler} field, this occurrence similarity offers evidence that the \emph{Kepler} sample may provide a good baseline for Galactic inferences. Furthermore, the \emph{Kepler} stellar sample is metal-rich compared to the \emph{K2} Campaign 5 sample, thus a finding of occurrence parity may reduce the role of metallicity in planet formation. However, a weak ($1.5\sigma$) difference, in agreement with metal-driven formation, is found when assuming the \emph{Kepler} model power-laws for the \emph{K2} Campaign 5 sample and optimizing only the planet occurrence factor. This weak trend indicates further investigation of metallicity dependent occurrence is warranted once a larger sample of uniformly vetted \emph{K2} planet candidates are made available. 

\end{abstract}



\section{Introduction}
The \emph{Kepler} mission continuously collected photometric data of over 150,000 stars for $\sim$3.5 years \citep{koc10,bor16}. This data set has provided evidence for nearly $4,500$ transiting exoplanet candidates.\footnote{\href{https://exoplanetarchive.ipac.caltech.edu}{https://exoplanetarchive.ipac.caltech.edu}} Most of the candidates were detected by the fully automated \emph{Kepler} pipeline \citep{jen10} and the {\tt Robovetter} \citep{tho18}, an automated vetting software built to distinguish real signals from false positives. Removing the human component of planet detection, the rate of false negatives (completeness) could be calculated with artificial transit injections \citep{pet13b,chr15,chr17}. The DR25 team also estimated the rate of false positives (reliability) by inverting the light curves, eliminating any real transit signals, and testing the ability of the pipeline to remove false detections \citep{cou17b}. With this set of planet candidates detected autonomously and an associated measure of the sample completeness and reliability, this data set has become the gold standard for deriving planet occurrence rates. 

\begin{figure*}
\centering \hfill \includegraphics[height=5.4cm]{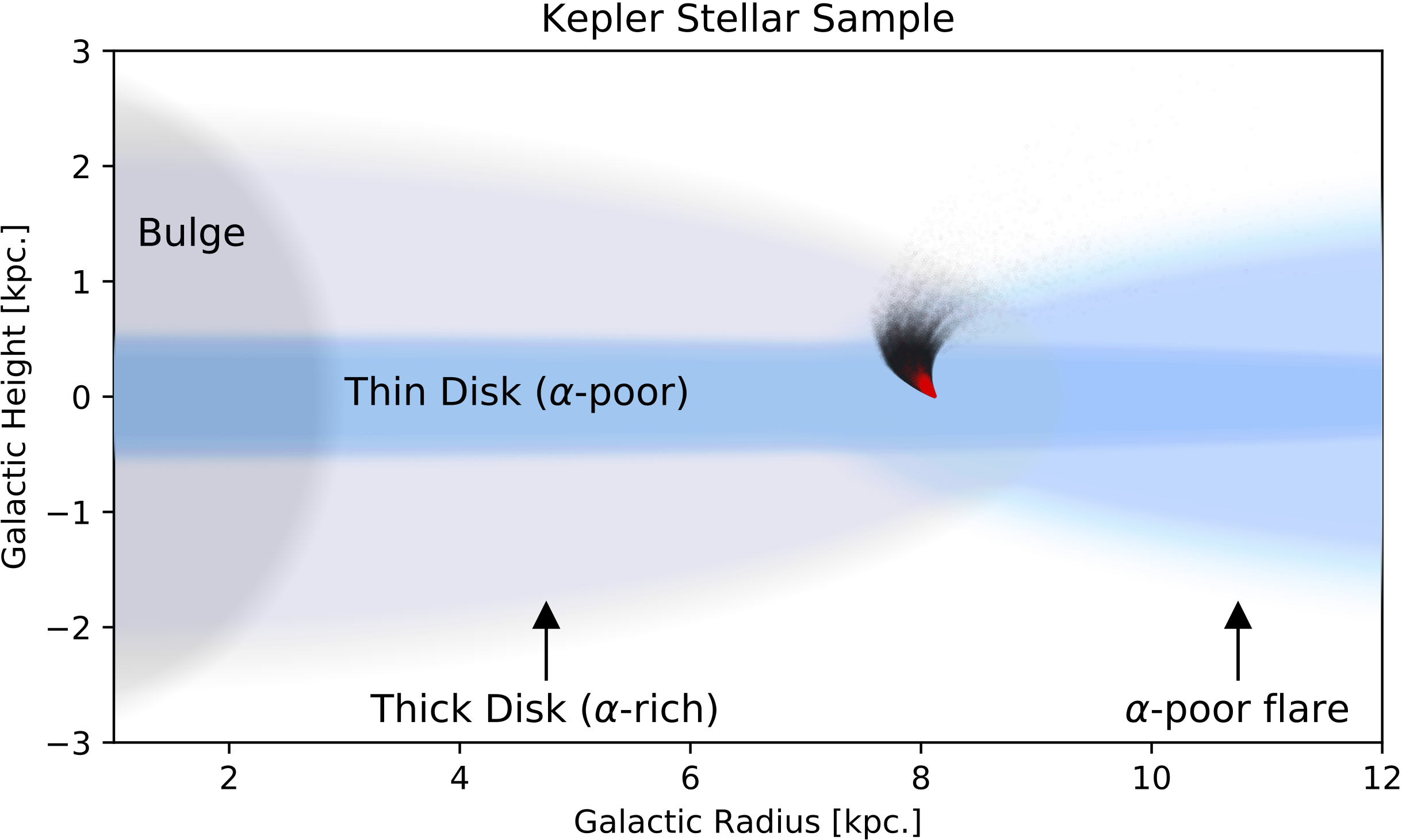}
\centering \hfill \includegraphics[height=5.4cm]{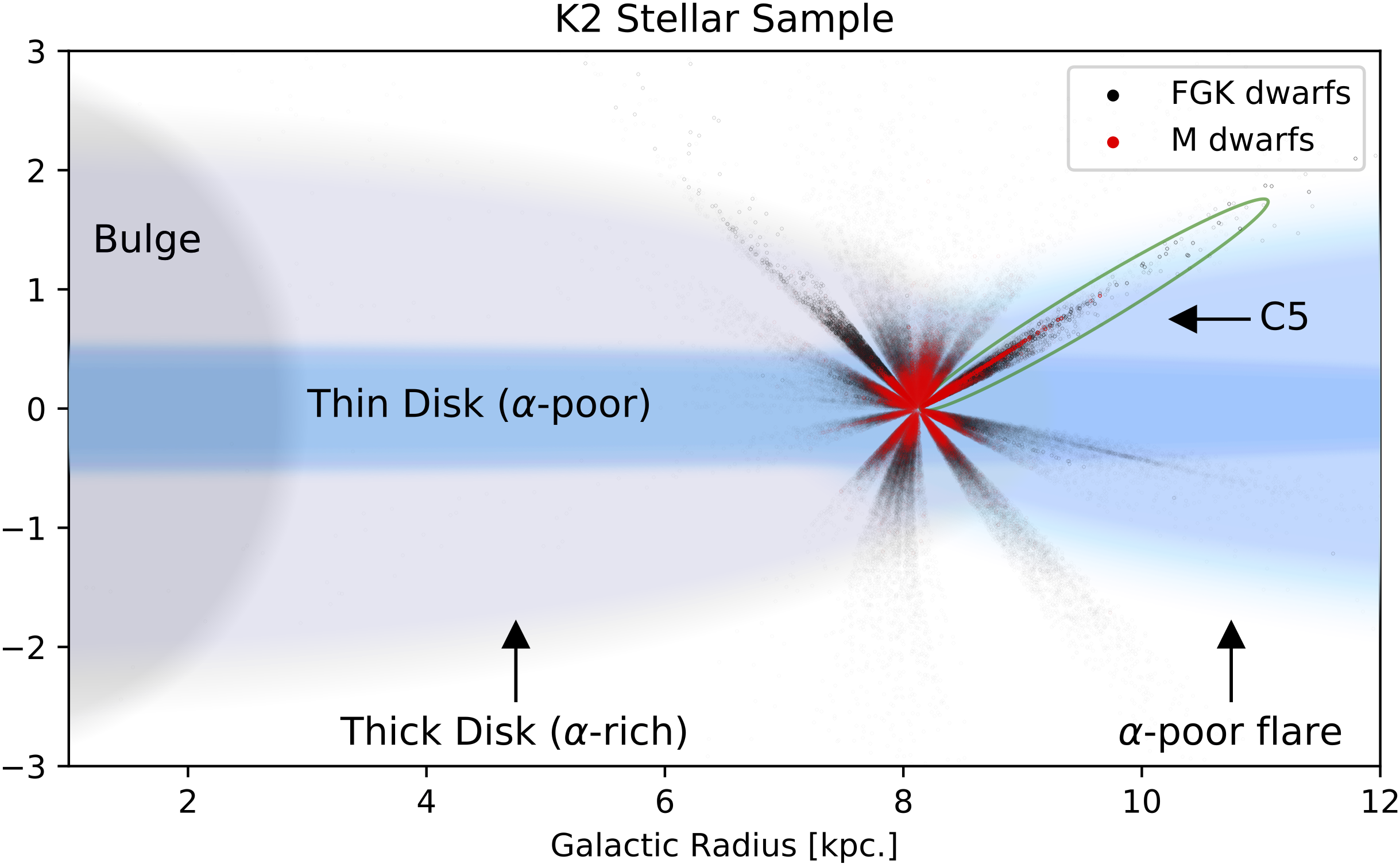}
\caption{The position of the M dwarf and FGK dwarf population as a function of the Galactic radius and height for both the \emph{Kepler} (left) and \emph{K2} (right) stellar sample (spectral classification was established using the \citet{ber20} and \citet{har20} stellar sample). The Galactic disk structure presented here follows the \cite{hay15} interpretation. With respect to iron, the $\alpha$-chain elements (O, Ne, Mg, Si, S, Ar, Ca, and Ti) exist in higher abundance in the thick disk (which truncates near the solar neighborhood) and lower abundance in the thin disk. Additionally, the $\alpha$-poor disk begins to flare up beyond the solar neighborhood. The \emph{K2} Campaign 5 stars, which are the focus of the current study, have been circled in green. The Galactic locations presented here are calculated using the \emph{Gaia} DR2 \citep{gai18}. 
\label{fig:galax}}
\end{figure*}

Numerous studies have used completeness and reliability to calculate planet occurrence rates for FGK dwarfs (i.e. \citealt{you11,how12,pet13b,muld18,zin19,he19}) and M dwarfs (i.e. \citealt{dres13,dres15,har19}). However, one critique of these studies is that they only sample a single 116 square degree patch of the sky --these occurrence rates may be specific to this region of the galaxy. Planet occurrence may change at Galactic latitudes with differing stellar metallicities, masses, radii, multiplicities, and stellar age, potentially highlighting the effects of stellar environment on planet formation. Additionally, these differences will modify the inferred Galactic exoplanet population. This point becomes more relevant when considering the calculated values of $\eta_{\Earth}$ \citep{cat11,tra12,pet13b,sil15,zin19b}, the probability of a potentially habitable planet being found around a given Sun-like star. Since all current estimates have used the \emph{Kepler} sample to extrapolate this probability, it remains unclear how well the value represents the census of habitable planets throughout the Galaxy.

Fortunately, the \emph{K2} mission (which came into existence after the \emph{Kepler} telescope lost functionality of two reaction wheels to stabilize the spacecraft) collected data from 18 different regions (Campaigns) across the Ecliptic plane \citep{how14,van16}. In Figure \ref{fig:galax} we show how the \emph{K2} sample probes different parts of the Galactic disk structure. A majority of the \emph{Kepler} sample is contained in the thin disk, which is $\alpha$-poor and consists of comparatively younger stars. The \emph{K2} sample probes deeper into the thick disk (older stars that are $\alpha$-rich; \citealt{sha19}), providing potential insight into the effects of age and $\alpha$ element abundance on planet occurrence. Additionally, the \emph{K2} sample spans 7-10 kpc in Galactic radius, while the \emph{Kepler} sample is limited to a range of 7.5--8.5 kpc in Galactic radius. This expansion allows us to investigate planet occurrence around stars with varying Galactic radii.

\emph{K2} photometry is affected by considerably more systematic pointing issues than the \emph{Kepler} prime mission. Therefore, the automated pipeline built for \emph{Kepler} data was not able to be successfully applied to the \emph{K2} data set. Various studies, which involved manual vetting, have found nearly 800 planet candidates in the \emph{K2} data \citep{van16,bar16,ada16, cro16,pop16,dre17,pet18,liv18,may18,ye18,kru19,zin19c}, but none have provided estimates of completeness and reliability due to the subjective nature of the searches. 

\begin{figure*}
\centering \includegraphics[height=7cm]{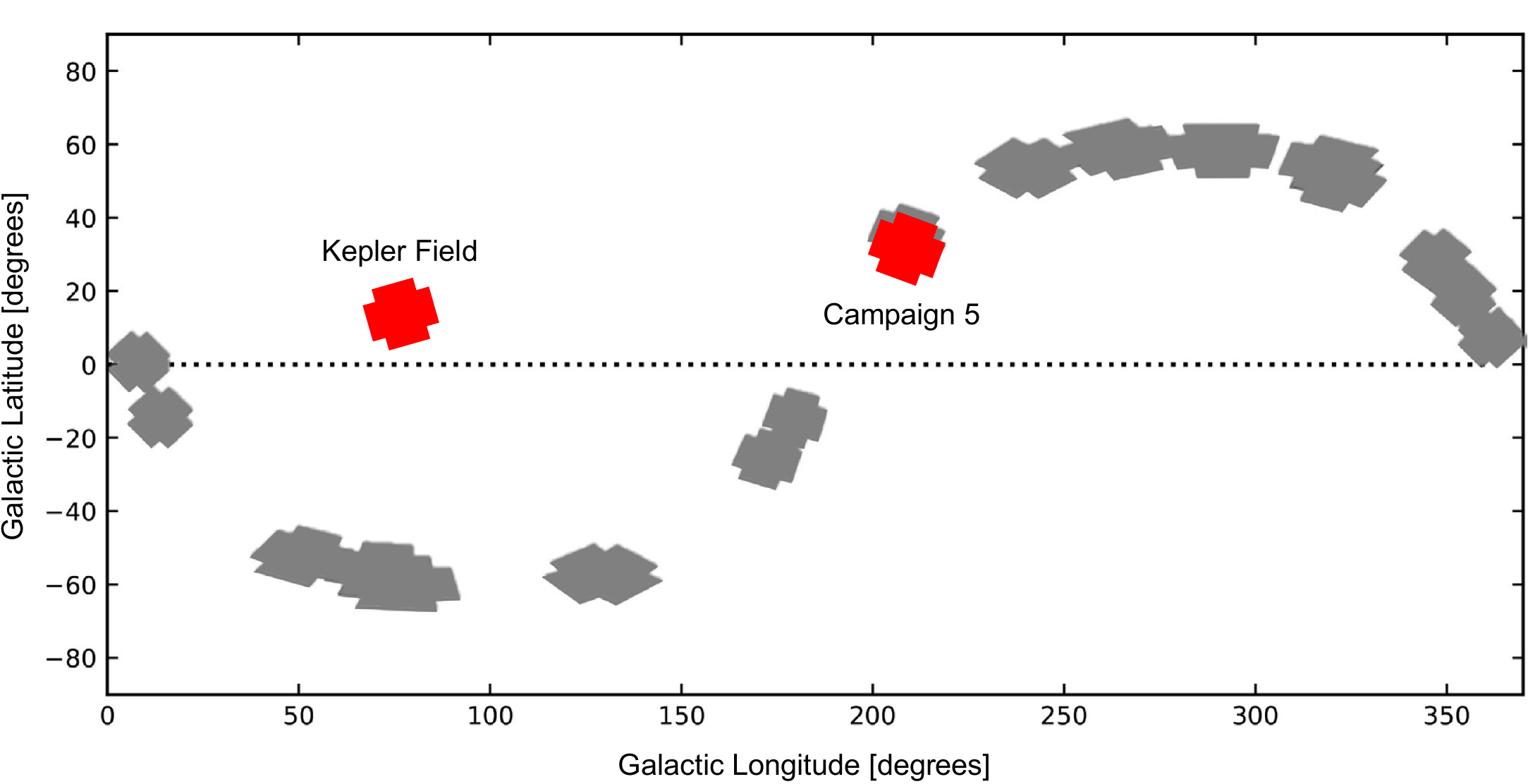}
\caption{The location of the 18 \emph{K2} fields are plotted in grey as a function of Galactic longitude and latitude. The \emph{Kepler} and Campaign 5 field, which are the subjects of this study, have been highlighted in red. These two fields are separated by about 120 angular degrees.
\label{fig:k2field}}
\end{figure*}

With the introduction of the {\tt EDI-Vetter} vetting software, \citet{zin20} were able to fully automate a planet detection pipeline for \emph{K2}. By removing the manual element from planet detection, \citet{zin20} were able to provide a uniform set of planet candidates with corresponding measures of completeness and reliability for the Campaign 5 field (henceforth, C5). In Figure \ref{fig:k2field} we show the separation between these two fields, illustrating the independence of these two samples. As uniform processing continues for the remaining 17 Campaigns, this early sample provides the first opportunity to consider small transiting planet occurrence outside of the \emph{Kepler} field.

In this paper we model the underlying exoplanet population for the \emph{K2} C5 FGK stellar sample using the forward modeling software {\tt ExoMult} \citep{zin19}. We provide discussion of our stellar and planet parameters and the cuts we make to isolate the samples of interest in Section \ref{sec:sample}. The forward modeling methodology is presented in Section \ref{sec:forward} along with modifications to {\tt ExoMult}. In Section \ref{sec:Result} we provide the optimized population models and discuss the occurrence parameters derived for each model. In Section \ref{sec:radgap} we consider the implication of the radius gap and other empirical features on our ability to model the population. We comment on the implications of our findings with regard to the effect of stellar metallicity in Section \ref{sec:metal}, and provide concluding remarks of this study in Section \ref{sec:conclusion}.

\section{Sample Selection}
\label{sec:sample}
In this section we discuss the stellar and planetary parameters used in this study. We also present the cuts made in both samples to ensure purity.

\subsection{\emph{K2} Stellar Selection}
\label{sec:k2Star}
\begin{figure}
\centering \includegraphics[width=8.5cm]{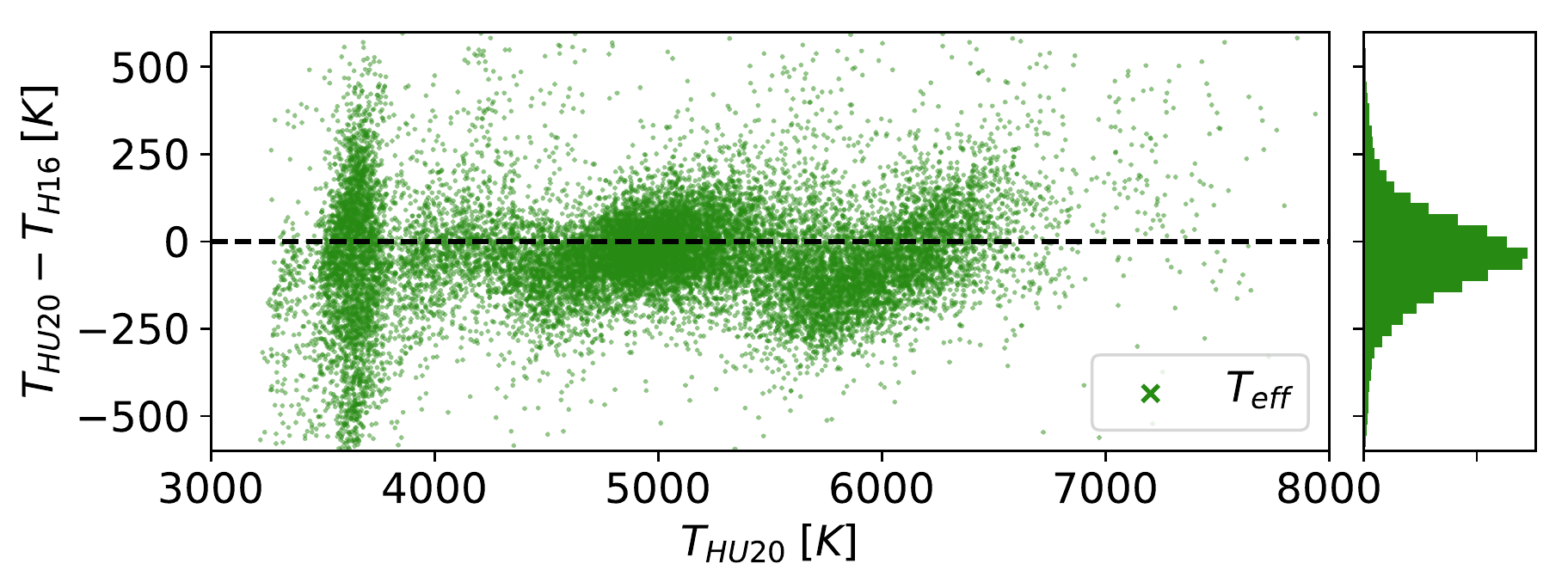}
\centering \includegraphics[width=8.5cm]{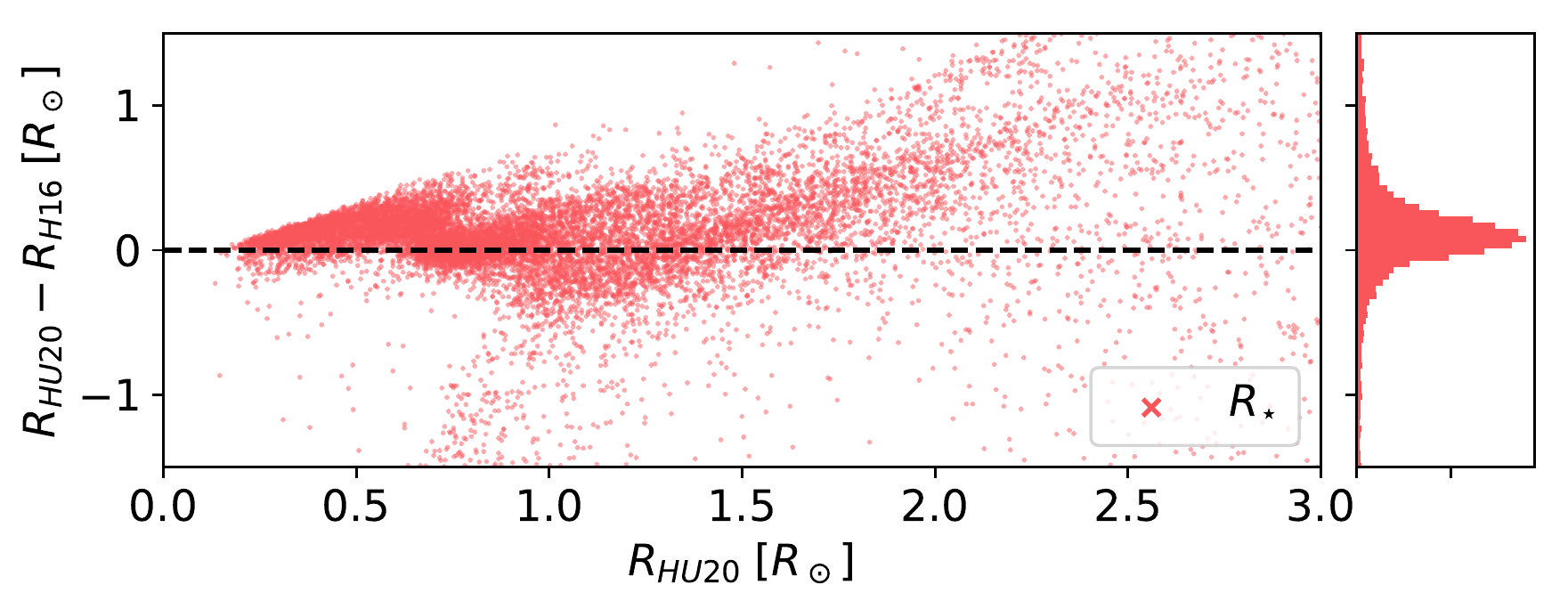}
\caption{Shows the overall change from H16 \citep{hub16} to HU20 \citep{har20}. The 19,004 C5 targets that overlap both catalogs are compared here. \textbf{Top:} We present the overall change in effective temperature ($T_{\textrm{eff}}$). We do find minor variations at different temperatures due to changes in spectral classification, but the overall systematic offset is not significant. \textbf{Bottom:} We present the overall change in stellar radius ($R_\star$) and find a very minor trend for small radius stars, but these M dwarfs are not included in our stellar sample. Overall the systematic offset is negligible. 
\label{fig:stell}}
\end{figure}

A key part of any planet occurrence rate is understanding the underlying sample of stars from which the detected planets are drawn. This can be accomplished by ensuring the stellar attribute measurements are as accurate as possible. If a significant portion of the sample parameters are inaccurate, we will under- or over-estimate the difficulty of detecting planets, and therefore bias the occurrence measurements. To minimize this effect we use the stellar parameters provided by \citet{har20}, as this provides the most uniform and up-to-date stellar parameters for \emph{K2}. Using \emph{Gaia} DR2 and \emph{LAMOST} spectroscopic data, this sample significantly reduces the uncertainty in stellar radius measurements compared to measurements derived primarily by photometry \citep{hub16}. However, due to the requirement of a \emph{Gaia} DR2 parallax, this catalog of random-forest derived parameters is only available for 19,220 stars out of the 25,030 potential targets studied by the \emph{K2} pipeline \citep{zin20}. In Figure \ref{fig:stell} we show the overall change in stellar radius and effective temperature between the \cite{hub16} and \citet{har20} catalogs. We find a median decrease of 37K in stellar temperature and a median increase of $0.11R_\odot$ in stellar radius. While the overall offset between catalogs is minor, we acknowledge that some structure can been seen between parameters. Most noticeably, the temperature spread of M dwarfs. Accurate measurements of this parameter are difficult to achieve for M dwarfs, leading to this spread. Fortunately, these stars are not included in our sample (as this work focuses on FGK stars). Overall, the apparent systematic structure between catalogs is minor and within the uncertainty of the parameter measurements. Since this effect is small, and most targets experience a minor change in parameter values between catalogs, we rely on the \cite{hub16} stellar parameters with photometrically derived metallicities for the remaining 5,810 stars. By including these additional stars in our sample, we are able to maximize the planet sample. However, we recognize the increased uncertainty introduced by including portions of the \cite{hub16} sample.  

It is important that all the stars in our sample have measured values of stellar temperature ($T_{\textrm{eff}}$), radius, mass, surface gravity (log(g)), and metallicity ([Fe/H]). The requirement of metallicity helps constrain the limb darkening parameters. Thus, we remove 3,020 targets that do not have these measurements available, reducing the sample to 22,010 stars. In all cases we find the target either has a measure of all five of these attributes or none of them.

Our second cut to the sample is meant to eliminate evolved giant stars. We implement the log(g) threshold derived by \cite{hub16}:

\begin{equation}
\textrm{log(g)} \ge \frac{1}{4.671} \textrm{arctan} \left( \frac{T_{\textrm{eff}}-6300\textrm{K}}{-67.172\textrm{K}} \right)+ 3.876
\label{eq:logg}
\end{equation}
which approximates the limit for dwarf classification (according to the Parsec models; \citealt{bre12}) for stars with solar-metallicity. Undoubtedly some of the stars in our sample deviate from solar-metallicity, but this threshold remains sufficient in eliminating evolved stars (see Figure 6 of \citealt{hub16}). For reference, this equation permits log(g)$\ge 4.20, 4.19$, and 3.61 for $T_{\textrm{eff}}=$ 4500K, 5500K, and 6500K respectively. We remove 6,895 stars that do not meet this requirement, leaving 15,115 stars in our stellar sample.

\begin{figure}
\centering \includegraphics[width=8.5cm]{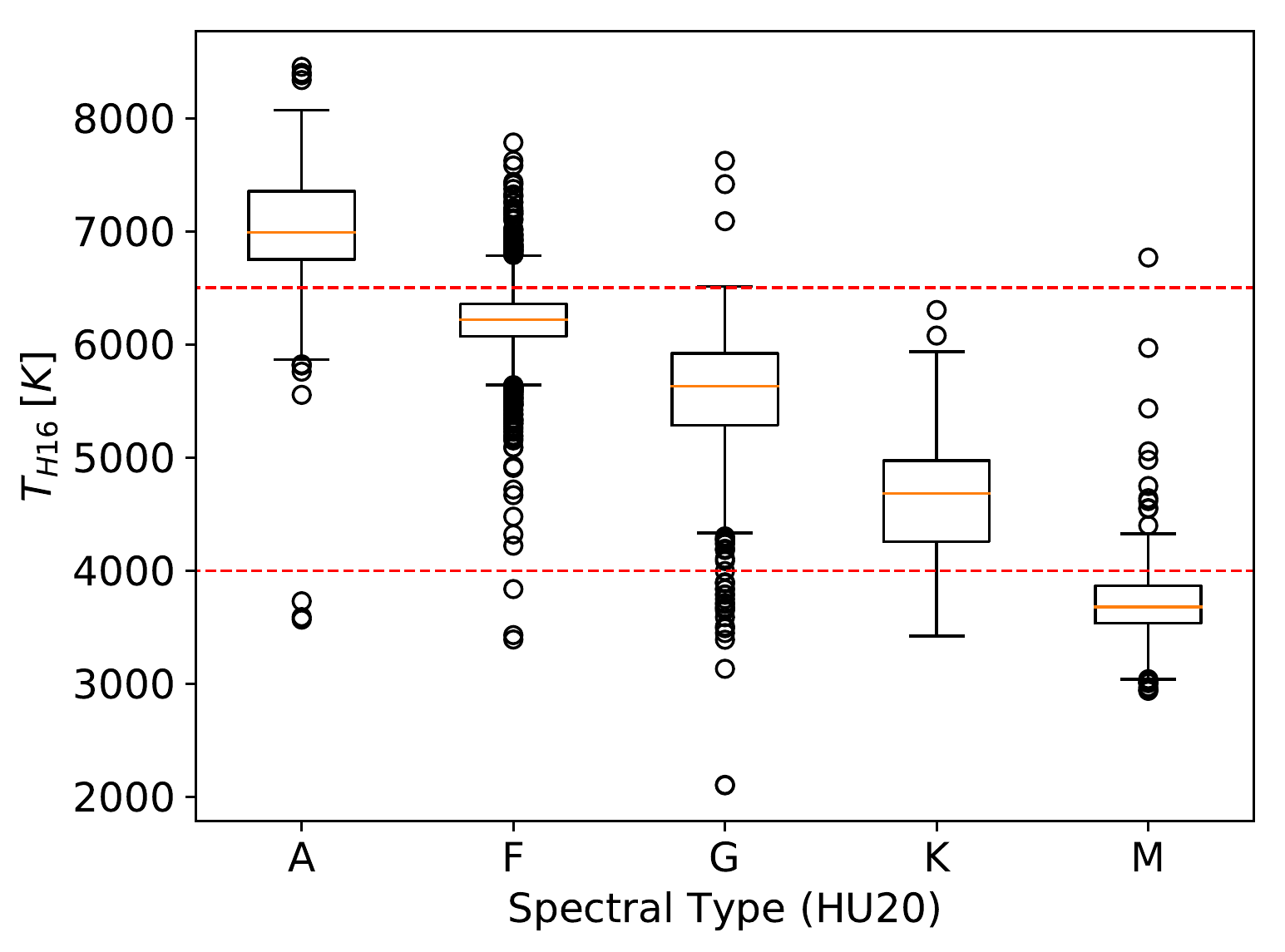}

\caption{Shows the inferred stellar temperature ($T_{\textrm{eff}}$) from H16 \citep{hub16} vs. the spectral classifications of HU20 \citep{har20}. The 13,029 C5 targets that overlap both catalogs and meet the log(g) threshold in Equation \ref{eq:logg} are compared here. The orange line represents the 50th percentile (median) of each set of values. The black box shows the 25th and 75th percentile of the data and the whiskers represented the lower and upper bounds of the data (excluding outliers which are displayed as flier points). The red dotted lines represent the limits selected for FGK classification from stars with only H16 parameters available.
\label{fig:spect}}
\end{figure}

The \citet{har20} catalog is unique in that it provides an estimate of the spectral type of the star. We utilize this feature to select for F, G, or K spectral types. Unfortunately, the \cite{hub16} does not provide a similar spectral measure, thus we rely on the inferred stellar temperature ($T_{\textrm{eff}}$). We consider a star in the FGK regime if $T_{\textrm{eff}}$ is within the range of 4000-6500K. In Figure \ref{fig:spect} we compare targets that overlap both catalogs and find this range best represents FGK classification in the \citet{har20} catalog. After removing 5,060 targets (4,808 M dwarfs and 252 A dwarfs), this cut allows 10,055 stars to remain in our sample. 

Our last filters remove stars that were deemed problematic by the \emph{K2} pipeline. These light curves were either unable to be properly smoothed, or the stellar surface is extremely active, making transit detection nearly impossible. We use the Combined Differential Photometric Precision (CDPP) to determine the threshold of transit detection. CDPP is a measure of the average noise found within the light curve, given a window of time \citep{chr12}. As suggested by \citet{zin20}, we remove targets with CDPP$_{8hr}$ > 1200ppm, eliminating 782 targets. Additionally, some targets had very large photometric apertures which likely contained more than one star. We exclude these targets by enforcing a maximum aperture threshold of 80 pixels, removing 16 stars. 

After refining our sample to meet the discussed requirements, we are left with 9,257 stars. Of these stars, only 865 rely on parameters from \cite{hub16}. The remaining 8,392 stars use parameters derived by \citet{har20}. Finally, we calculate the two quadratic limb darkening values for each target, using the ATLAS model coefficients for the \emph{Kepler} bandpasses tabulated by \citet{cla12}.

With our stellar sample in hand, we caution that \emph{K2} field selection was guest observer driven. This could potentially bias our sample to focus on regions with bulk stellar properties (i.e. mass, radius, and metallicity) that favor planet detection. Upon examining a $10\degr$ radius of the C5 field using the TIC (TESS Input Catalog)\footnote{\href{https://tess.mit.edu/science/tess-input-catalogue/}{https://tess.mit.edu/science/tess-input-catalogue/}}, we find the stellar parameter distributions of the C5 sample do not deviate from that of the broader field. We can therefore conclude that the guest observer selection effect will be negligible and disregard such issues in this study.     

\subsection{\emph{Kepler} Stellar Selection}
\label{sec:kepStar}
As the main objective of this paper is to compare \emph{Kepler} occurrence to that of the \emph{K2} C5, we also use similar cuts to select our \emph{Kepler} sample. We begin with the stellar parameters provided by the \citet{ber20} catalog (186,301 stars). By limiting the sample to stars that meet the log(g) requirements of Equation \ref{eq:logg}, a $T_{\textrm{eff}}$ within the range of 4000--6500K, a measured CDPP$_{7.5hr}<1000$ppm, and available measurements of radius and mass, we are left with 104,498 stars. Here, we have relaxed the requirement of metallicity for this sample and instead use Equation 9 of \citet{zin19}, which provides a method of determining the limb darkening parameters used by the \emph{Kepler} DR25 detection pipeline using only the star's $T_{\textrm{eff}}$. To ensure our sample only includes stars with significant data available, we remove targets with less than two years between the first and last photometric data points (span > 730 days) and only include targets with more than 60\% of the cadences available between these end points (duty > 0.6). This final cut leaves 94,222 stars in our stellar sample.  

\begin{figure}
\centering \includegraphics[width=8.5cm]{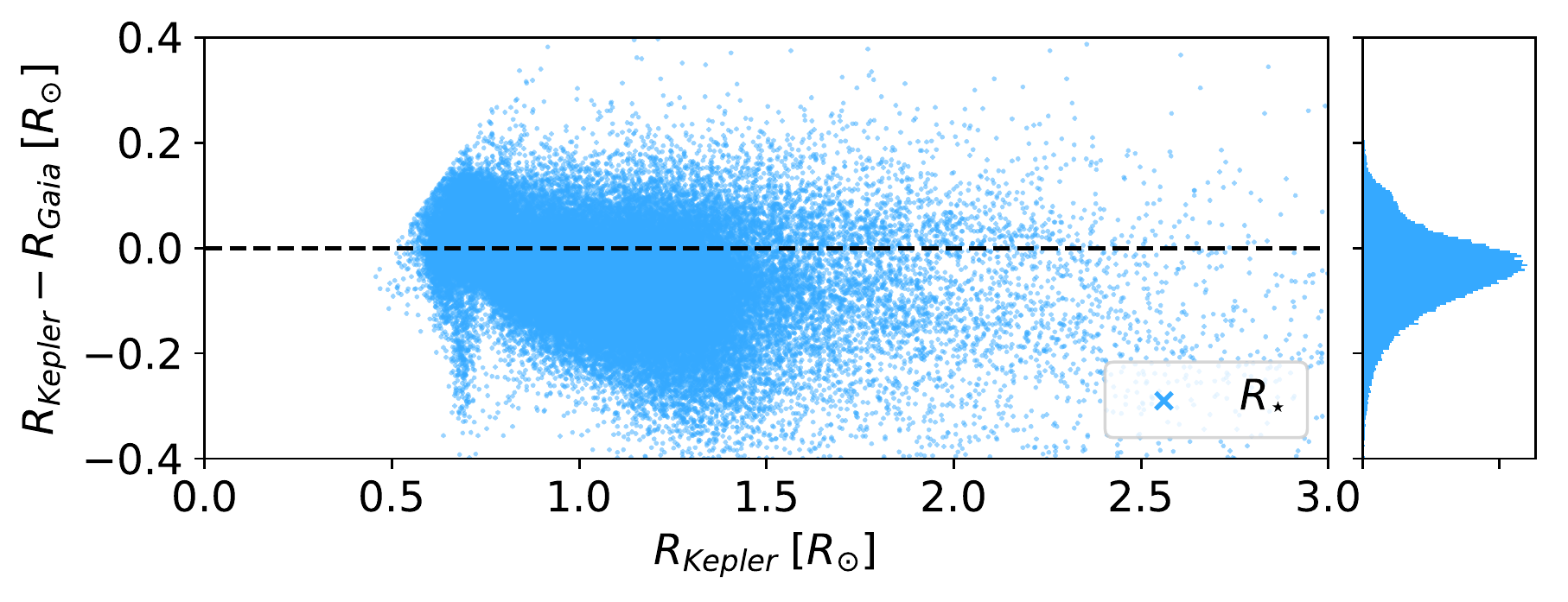}
\centering \includegraphics[width=8.5cm]{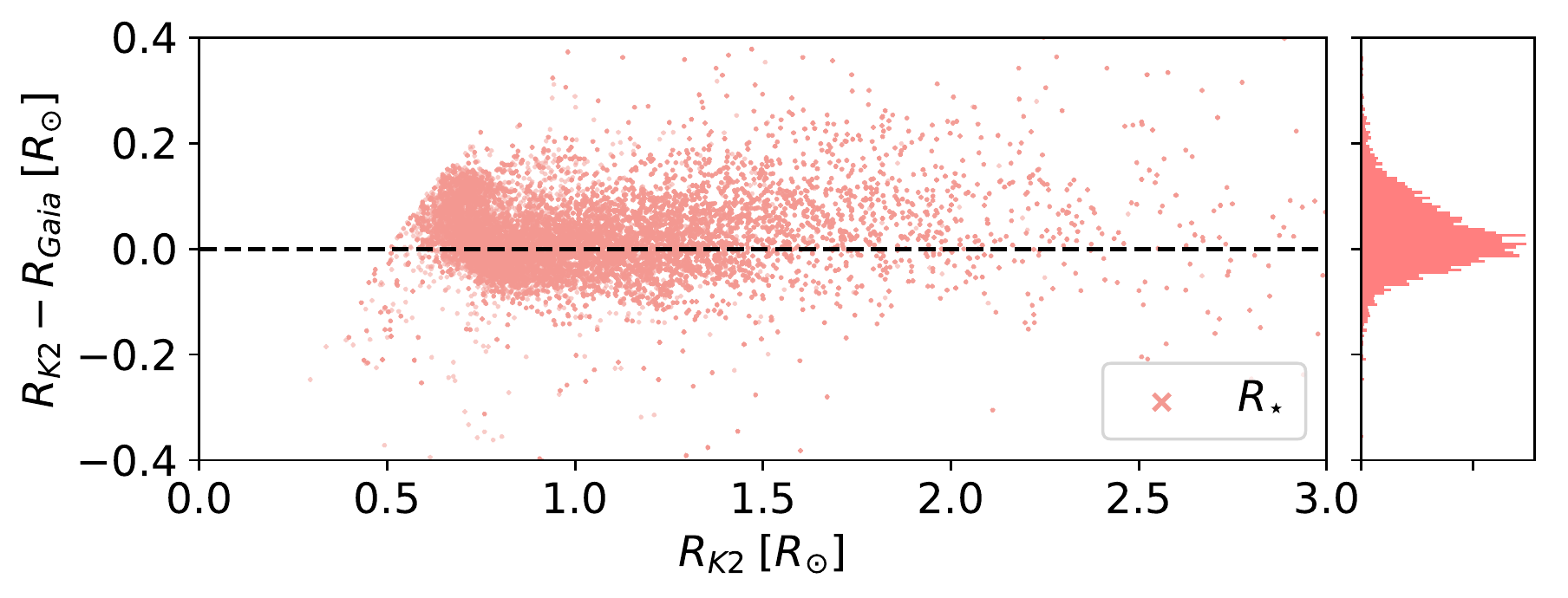}

\caption{A comparison of the radius values used by our \emph{Kepler} (\textbf{Top}) and \emph{K2} (\textbf{Bottom}) stellar catalogs with the radius values derived by the \emph{Gaia} team \citep{gai18}. Overall, both catalogs find similar values with those uniformly derived by \emph{Gaia}.
\label{fig:gaiaCom}}
\end{figure}

As noted for the \emph{K2} stellar sample, using catalogs with uniquely derived parameters can introduce potential systematic offsets. The stellar radius measurements are the most concerning, as these have the largest effect on the inferred planet occurrence. To address this issue, we compare the radius measurements of both our \emph{Kepler} and \emph{K2} samples to the radius values uniformly derived by the \emph{Gaia} team \citep{gai18}. In Figure \ref{fig:gaiaCom} we find a very minimal systematic offset between the \emph{K2}-\emph{Gaia} sample ($0.02R_{\Sun}$) and the \emph{Kepler}-\emph{Gaia} sample ($-0.04R_{\Sun}$), indicating that differences caused by unique parameter derivation will be relatively small. This finding is not surprising as both catalogs used \emph{Gaia} DR2 to infer their radius parameters. However, the additional photometry data used by both \citet{ber20} and \cite{har20}, to infer stellar parameters, provides increased accuracy when compared to that \emph{Gaia} team measurements. Therefore, we use our original \emph{K2} and \emph{Kepler} catalog values while acknowledging their minor systematic offsets.

\begin{figure*}
\centering \includegraphics[height=5.75cm]{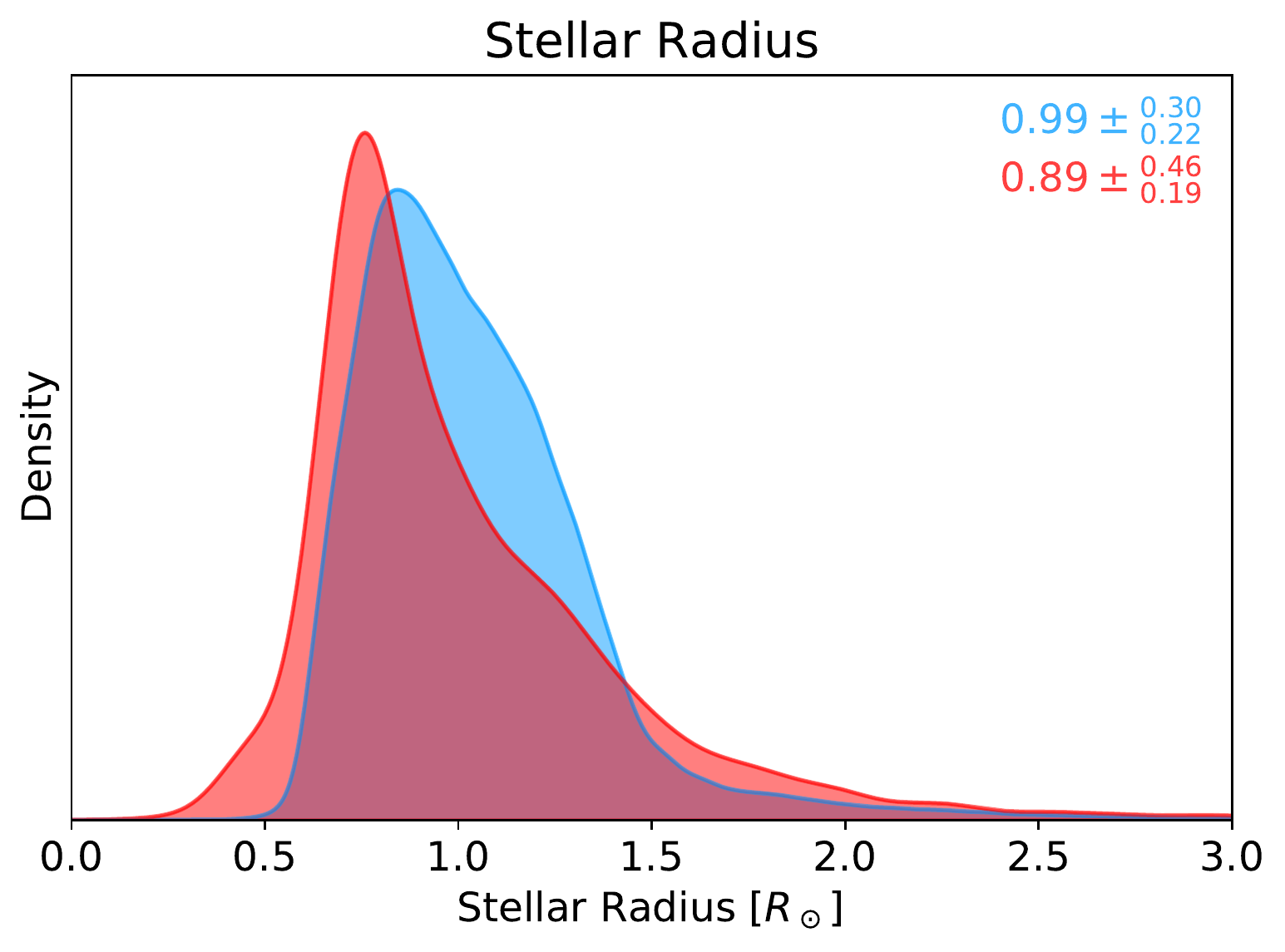}
\includegraphics[height=5.75cm]{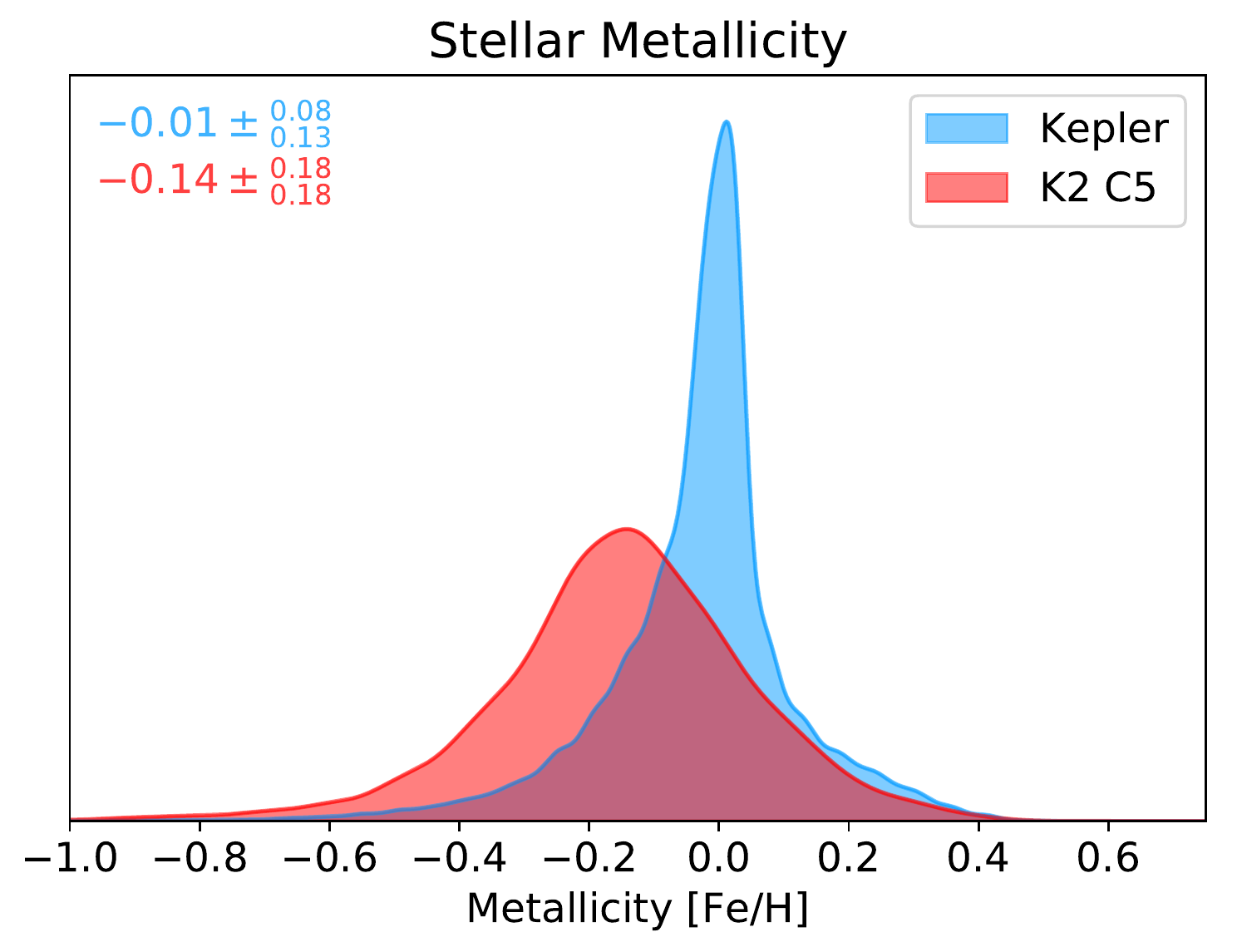}
\vspace{2 mm} \\
\centering \includegraphics[height=5.75cm]{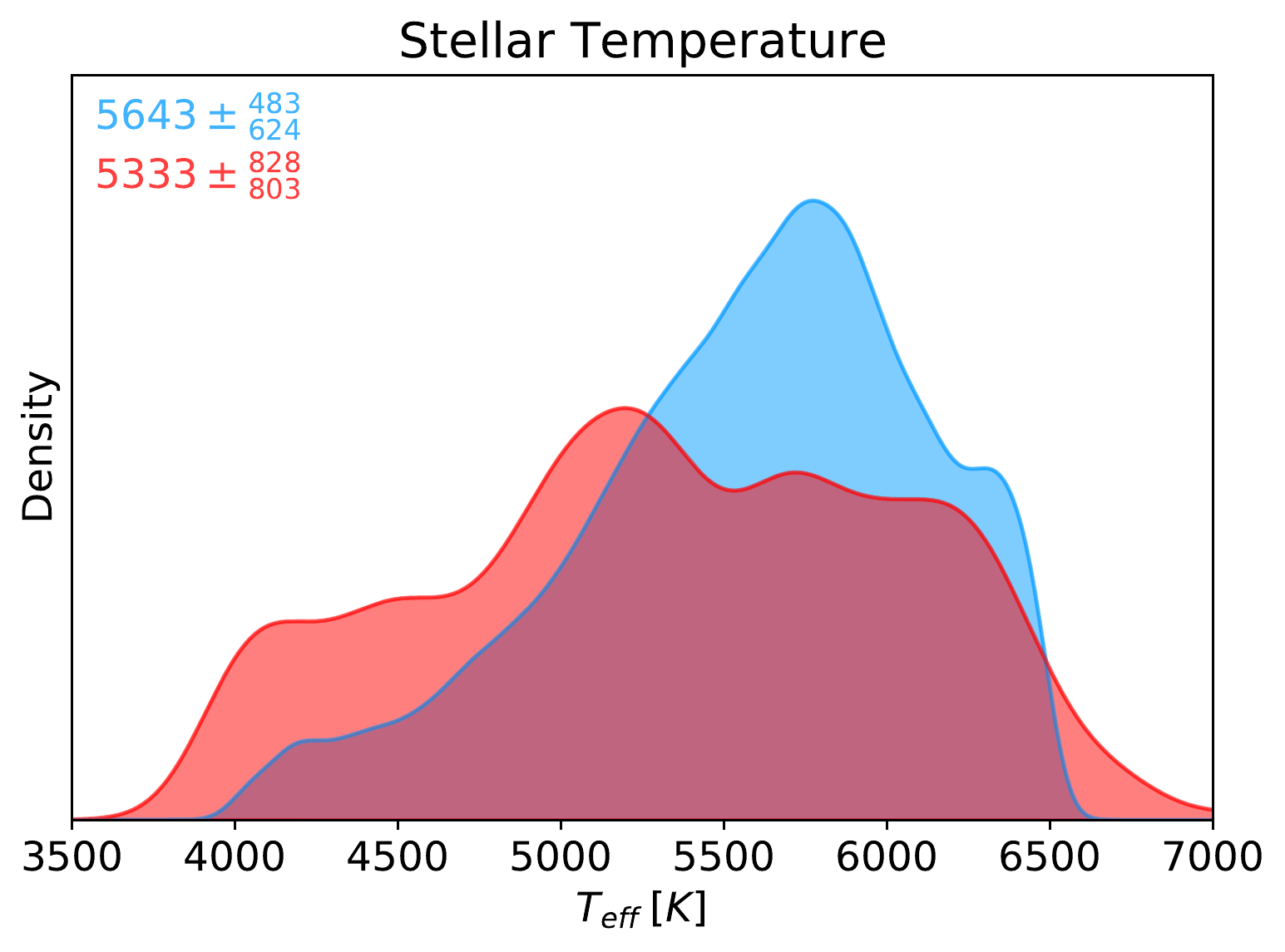}
\includegraphics[height=5.75cm]{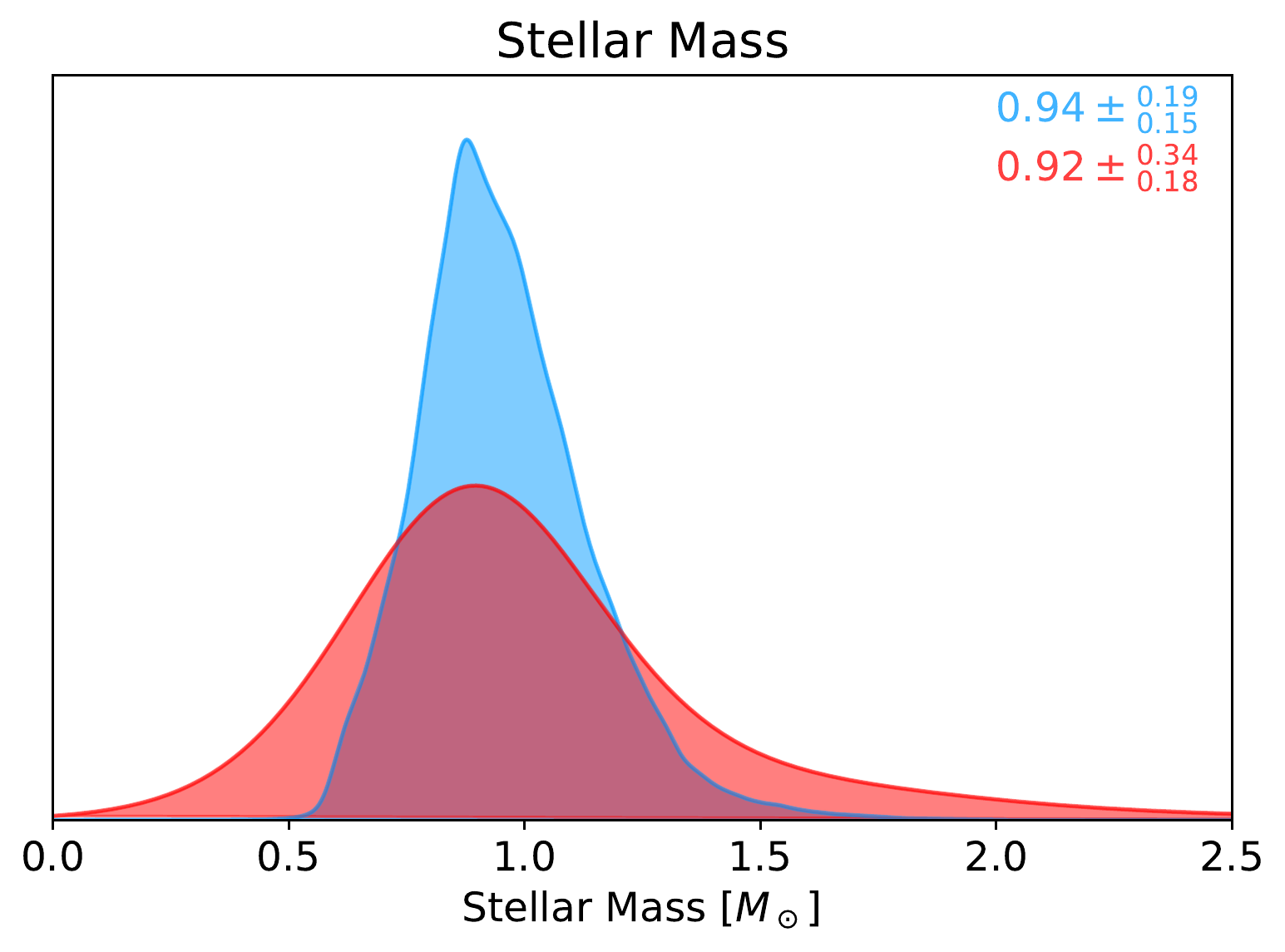}
\caption{Shows the distribution of stellar parameters for our \emph{Kepler} and \emph{K2} C5 FGK dwarf sample. Measurements of radius, mass, metallicity, and $T_{\textrm{eff}}$ use the parameter values provided by \citet{ber20} for \emph{Kepler} and \citet{har20} \& \citet{hub16} for \emph{K2} C5. The 50th, 16th and 84th percentiles have been listed in the upper corner of each plot.  
\label{fig:stellarCom}}
\end{figure*}

\subsection{\emph{K2} vs. \emph{Kepler} Stellar Sample}

The \emph{Kepler} field and the \emph{K2} C5 field represent independent samples in which we can measure planet occurrence in our local Galaxy. Furthermore, the stars in these samples are unique and provide insight into the stellar features that inhibit or encourage planet formation. In Figure \ref{fig:stellarCom} we compare the distributions of stellar parameters across both samples. Overall, the \emph{K2} C5 sample appears contains stars that have smaller radii, are slightly cooler, and are metal-poor by comparison.

Early evidence from radial velocity surveys found an increased hot Jupiter occurrence around metal-rich stars, suggesting metal-rich protoplanetary disks are able to form planets more efficiently \citep{fis05}. However, the \emph{Kepler} data provided evidence for a lower occurrence of hot Jupiters ($0.5\pm0.1\%$ of stars; \citealt{how12}) compared to the local solar neighborhood population ($1.20\pm0.38\%$ of stars; \citealt{wri12}). With data from LAMOST, \citet{don14} was able to show that the \emph{Kepler} field has a near-solar mean metallicity ([Fe/H]$=-0.04$ dex), which is comparatively higher than the local solar neighborhood ([Fe/H]$=-0.14\pm0.19$ dex; \citealt{nor04}). Additional evidence of this positive stellar metallicity offset was provided by \citet{guo17}, indicating that a metal deficiency cannot explain the reduced occurrence of hot Jupiters in the \emph{Kepler} sample. This discrepancy may lead one to minimize the role of metallicity in planet formation. However, an increased sub-Neptune population was found around metal-rich stars within the \emph{Kepler} sample \citep{pet18}, indicating the role of metallicity is more nuanced than previously believed. Clearly, more data are needed to parse out the details of this effect.

Remarkably, the \emph{K2} C5 sample provides a metallicity distribution that is very similar to solar neighborhood distribution ([Fe/H]$=-0.14\pm0.18$ dex). If an overall decrease in planet occurrence was found in this sample, it would provide additional evidence for a metallicity-dependent formation mechanism. In the absence of an occurrence deficiency, the role of metallicity remains nuanced and beyond the detection of our broad summary statistics. Once additional \emph{K2} Campaigns are available, detailed studies considering the effects of $\alpha$-chain element abundances can be accomplished with \emph{K2} data (see Figure \ref{fig:galax}) as these elements appear to be correlated with the detection of planets \citep{adi12}.

\subsection{Planet Selection}

\subsubsection{\emph{K2} C5 Planets}

In this study we use the \emph{K2} Campaign 5 planet sample obtained by \citet{zin20}. This catalog is a uniformly vetted sample with a corresponding measure of completeness and reliability. This sample includes 75 planet candidates that are at least 94.2\% reliable (small number statistics only allow for a lower limit on the measure of reliability). For our sample we adopt the planet parameters of radius and period derived in \citet{zin20}.

The planet sample is drawn from the subset of stars selected in Section \ref{sec:k2Star}. This cut removes 26 planets from our sample: 18 M dwarf candidates, one high CDPP light curve candidate, six low log(g) stellar host candidates, and one candidate without measured stellar parameters.

In addition, we remove gas giant planets from our sample. It has been shown these giants have a tendency to eject planets as they migrate inward \citep{bea12}. This inward orbital migration creates an independent population of giant planets that do not share the same population features as the planets formed in-situ (further evidence of this unique population was noted by \cite{joh12}, who showed that multi-planet systems with a short period planets greater than 0.1 Jupiter mass were dynamically unstable on short timescales). Additionally, it has been shown empirically that planets with $R>6.7R_{\Earth}$ form a unique population that deviates from a simple power-law \citep{ste12}. This cut of $R>6.7R_{\Earth}$ roughly corresponds to planets with a mass greater than 0.1 Jupiter mass, derived dynamically by \cite{joh12}. However, the existence of short period gas giants in multi-planet systems has been seen in the \emph{Kepler} data set (i.e Kepler-56; \citealt{hub13}) and the \emph{K2} data set (i.e. WASP-47; \citealt{bec15}). While rare, these multi-planet systems indicate some small fraction of these massive planets can co-exist with other planets. As done in \cite{zin19}, we only remove planets with $R>6.7R_{\Earth}$ if no other planets were detected in the system. This cut removes six hot Jupiters from our sample. No multi-planet systems with an $R>6.7R_{\Earth}$ planet exist in the \emph{K2} C5 sample used for this study, however to maintain consistency with our \emph{Kepler} sample, we allow such systems in our forward model. Cutting the sample in this manner introduces a mild bias, as some of the planets removed may be part of systems with undetected planets. Additionally, the measured planet radius value may differ from the true value, modifying the parameter location of the planet relative to this cut. We account for this bias in Section \ref{sec:exomult}.

Despite the improved radius measurements provided by \citet{har20}, planet radii are still uncertain to about 16\% for most planets in this catalog. We assume the values provided are accurate, but address the biases these uncertainties produce in our forward model (Section \ref{sec:exomult}).

Overall, our sample consists of 43 planets with radii ranging from 1.2--6.3$R_{\Earth}$ and orbital periods ranging from 1.54--35.40 days. In Figure \ref{fig:planetCom} we present our planet sample and the expected detection probability. Empirically, this sample contains 34 single planet systems, three double planet systems, and one triple planet system.

\subsubsection{\emph{Kepler} Planets}

\begin{figure*}
\centering \includegraphics[height=8cm]{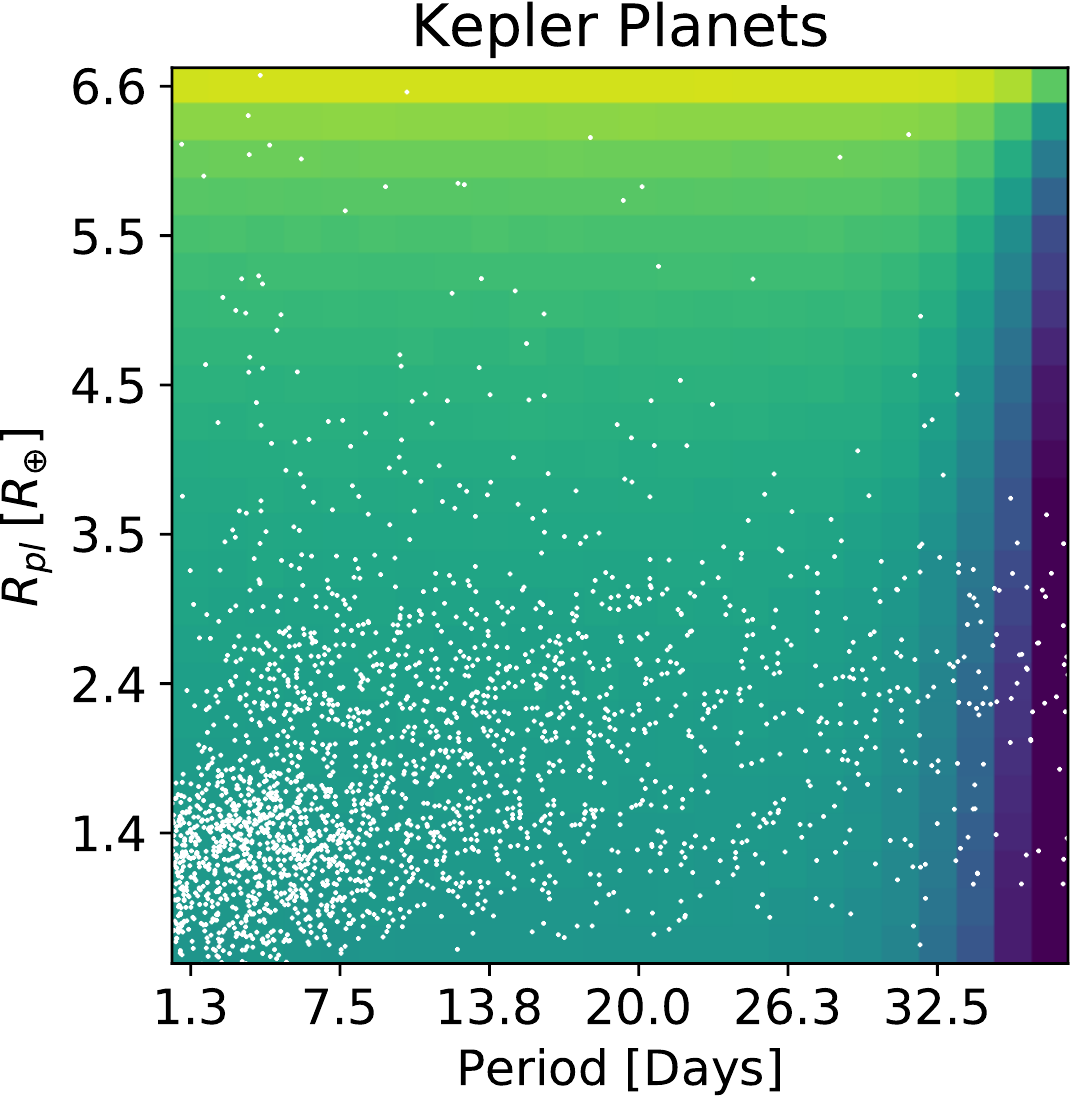}
\includegraphics[height=8cm]{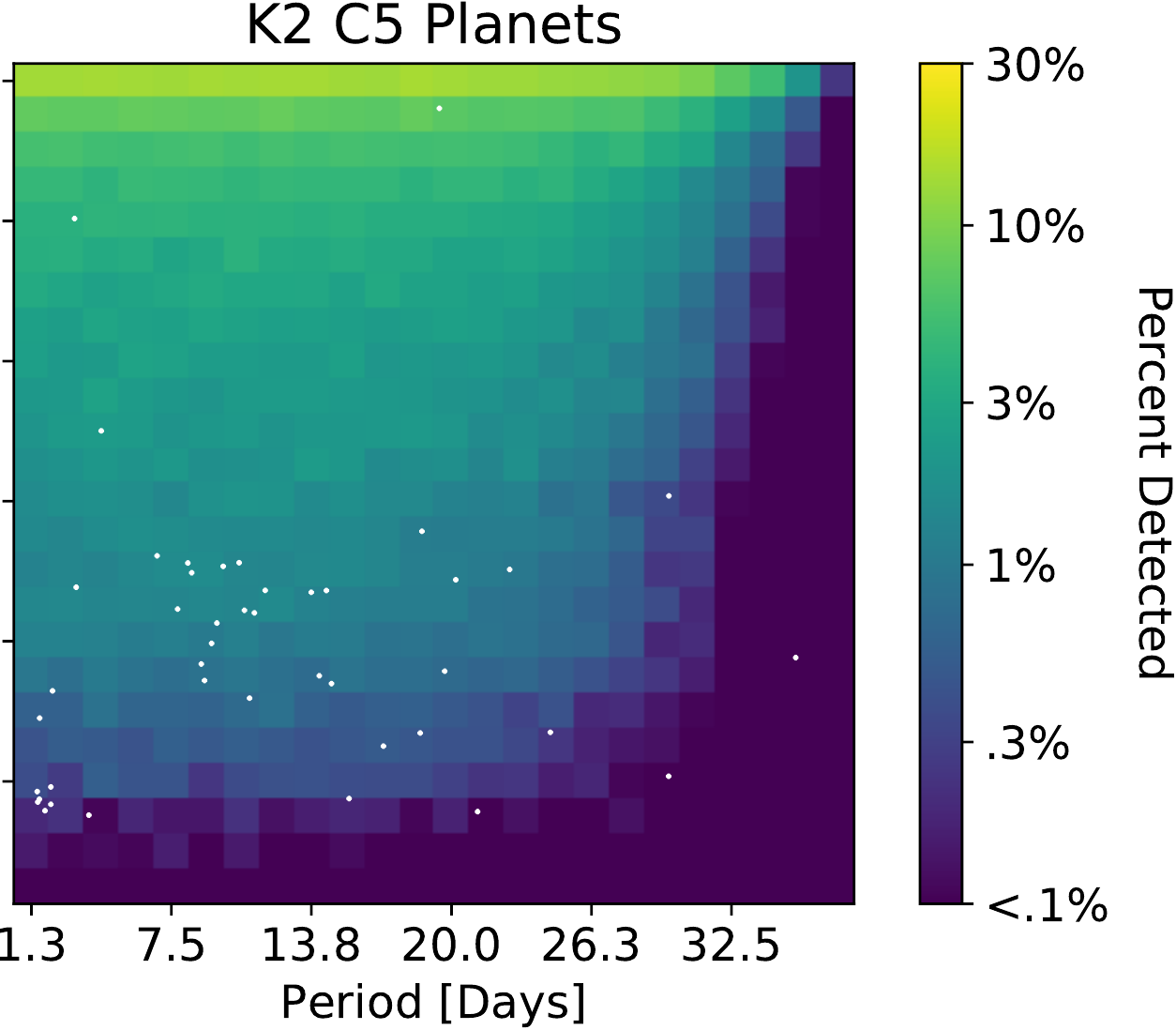}
 \caption{Shows the distribution of our planets samples in white for both \emph{Kepler} (\textbf{Left}) and \emph{K2} C5 (\textbf{Right}). Beneath this plot is a detection probability map. Using the stellar sample from each field a single planet, drawn uniformly over each bin of radius and period, is tested for detection through {\tt ExoMul}. The fraction of planets recovered is then portrayed in this color map.  
\label{fig:planetCom}}
\end{figure*}

We use the \emph{Kepler} planet candidate parameters provided by \citet[][4,612 planets]{tho18} with radius updates from \citet{ber20b}. Drawing planets from the stars selected in Section \ref{sec:kepStar} and applying the same gas giant removal procedure performed on the \emph{K2} C5 sample, we are left with 3,023 planets. This ensures the samples are comparable. The only difference is that we remove \emph{Kepler} planets with periods greater than 38 days (2,318 planets remain), as this is the longest detectable period for the \emph{K2} C5 data. In Figure \ref{fig:planetCom} we present our planet sample and the underlying sample completeness.

Overall, our sample contains 2,318 planets with periods ranging from 0.51--38.00 days and radii ranging from 0.50--11.45$R_{\Earth}$. This sample has an average reliability of 98.3\%, using the values provided by \citet{tho18}. Only 8 planets exceed the $6.7R_{\Earth}$ single planet radius cut. Again these large multi-planet systems are rare and have little effect on the overall occurrence measurements (as the radius power law decays quickly in this region of parameter space, $\sim R^{-4}$). However their inclusion allows us to account for fluctuations in and out of sample near the $6.7R_{\Earth}$ boundary due to measurement uncertainty. 

\section{Forward Model}
\label{sec:forward}
We use the {\tt ExoMult} software to forward model our population \citep{zin19}. This program takes a population of planets and subjects them to the selection effects of the detection pipeline, providing an expectation for the observed population. {\tt ExoMult} was originally designed to address the issues of the \emph{Kepler} pipeline \citep{tho18}, but the \emph{K2} data set has unique issues which require modification to the base code. We discuss these differences below.

\subsection{ExoMult}
\label{sec:exomult}

The original {\tt ExoMult} code assumes the planet radius and period distributions are independent, and modeled by broken power laws. Here, we adopt these same assumptions. One of the goals of this study is to make a comparative statement about the \emph{K2} sample versus the \emph{Kepler} sample. Trimming the \emph{Kepler} sample to match our 38 day period limit accomplishes this comparative goal, but we must carefully consider how doing so affects the optimization. A simple cut and re-process could yield inaccurate values, as non-detections provide constraints when optimizing the model. To avoid such issues we remove the window function from our \emph{Kepler} forward model. This function determines the probability of at least three transits appearing in the data. Since the \emph{Kepler} data set spans roughly 3.5 years, all of the planets within a 38 day period range would have experienced more than three transits, making this function unnecessary. In contrast, the \emph{K2} C5 data set spans roughly 75 days. This means that the three transit window function will be important for optimization. \citet{zin20} provides window function data for each light curve, but implementing these data into our forward model is computationally expensive. Instead we use the theoretical window function provided by \citet{zin20}, which closely matches the expected window function of most targets. 

Several previous studies have empirically found an average mutual inclination around 1--2$\degr$ \citep{lis11,lis12,fan12, muld19}. Implementing these non-independent details into our forward model is straight forward and a real advantage to this method of occurrence calculation. However, doing so requires a few additional parameters. This is not problematic for the \emph{Kepler} sample with greater than 2000 data points, but begins to verge on over-fitting when attempted on the 43 data point sample of \emph{K2} C5. For simplicity, we aim to minimize the number of variable parameters in this study. Therefore, we assume a mutual inclination of $0\degr$ and that the planets all orbit in a flat disk. Additionally, we assume all the planets exist on a circular orbit. This assumption is reasonable because nearly all of the planets with orbital periods $<38$ days will have experienced tidal circularization, resulting in a population with eccentricities near zero. 

{\tt ExoMult} was built to mimic the selection effects of the \emph{Kepler} pipeline. However, the \citet{zin20} planet catalog is far less complete than the \emph{Kepler} sample (See Figure 7 of \citealt{zin20}). Thus, we adopt the vetted completeness function of \citet{zin20} for our \emph{K2} C5 processing. For \emph{Kepler}, \citet{zin19} showed that additional signals in the same light curve had a lower detection efficiency. This effect has not yet been measured for the \emph{K2} sample. In an effort to make this a fair comparison between the two samples, we turn off this additional multiplicity completeness accounting in the \emph{Kepler} forward model optimization. 

One new feature introduced to {\tt ExoMult} is the ability to deal with radius uncertainties. Previous studies have addressed this issue using hierarchical Bayesian analysis \citep{for14,hsu19}, but forward modeling provided a straight forward method of accounting for these fluctuations. Each planet population is sampled and subject to all the selection effects of the \emph{Kepler} or \emph{K2} pipeline accordingly. The true stellar radius ($R_{\star}$) and planet radius ($R_{p}$) values are used to calculate the expected depth of the transit ($TD=R_{p}^2/R_{\star}^2$). To mimic radius variations caused by inaccurate depth measurements, we draw the measured transit depth ($\overline{TD}$) from a Gaussian distribution centered around the expected depth and a width of 4\% the expected depth (the median depth uncertainty determined by \citealt{zin20}). Independently, we draw the measured stellar radius ($\overline{R_{\star}}$) from a split normal distribution centered around the true radius value with a spread reflecting the upper and lower radius uncertainty. Using our drawn depth and drawn stellar radius values, we calculate the measured planet radius ($\overline{R_{p}}$):

\begin{equation}
\overline{R_{p}}=\sqrt{\overline{TD}}*\overline{R_{\star}}  . 
\end{equation}
The $\overline{TD}$ value is drawn independently for each planet in a given system, while $\overline{R_{\star}}$ is only drawn once per system. This method of drawing measured values allows us to account for fluctuations introduced by poor radius measurements. To address the bias introduced by our giant planet removal, we remove single planet detection systems with $R>6.7R_{\Earth}$ after the uncertainty modification has been applied.

\subsection{MCMC}
Using the expected populations provided by {\tt ExoMult}, we can optimize the model to produce an observed population similar to that of our sample. Under the assumptions listed in Section \ref{sec:exomult}, we have seven model parameters. The six broken power-law parameters (three for radius and three for period), and one overall occurrence factor ($f$). This $f$ factor tells the forward model what fraction of systems have a planet. In previous versions of {\tt ExoMult}, this factor was broken up to account for the number of stars with a planet and then the multiplicity of these systems. However, we want to minimize the number of parameters in this optimization to avoid over-fitting the \emph{K2} data. Thus, the $f$ parameter represents an overall measure of planet occurrence in these samples. If $f$ is greater than one, each star will be assigned a planet, and the excess fraction will be assigned a second planet. For a detailed discussion of the model parameters we refer the reader to Section 7 of \cite{zin19}. We measure the Bayesian posterior for the seven model parameters using the {\tt emcee} affine invariant sampler \citep{goo10,for13} with 50 semi-independent walkers, 5000 burn in steps, and 10,000 sample steps (500,000 total samples of each posterior). 

\subsection{Priors}
One of the features of Bayes theorem is the ability input prior information about your model parameters. For our \emph{Kepler} model and our \emph{K2} model, we assume uniform priors for all parameters. In order to avoid nonphysical cases, we allow $f$ to range from 0--7. All power law parameters are allowed to range from -20 to +20, and the radius break and period break are allowed to range from 0--16$R_{\Earth}$ and 0--35 days, respectively. While these uniform priors have little effect on our current fitting, in Section \ref{sec:K2/Kepler} we discuss how priors can be used to combine \emph{K2} and \emph{Kepler} samples in future studies.

\subsection{Likelihood Function}
To compute the likelihood function of our model we utilize two test statistics. First, we utilize the K-sample Anderson-Darling test statistic ($AD$; \citealp{and52,pet76}) to capture the shape of the distribution. Second, we use a marginalized Poisson distribution to ensure our inferred distribution is properly normalized.

Our shape metric uses the $AD$ test, a non-parametric method of measuring the probability that two samples come from the same distribution. In our case the two samples are the observed \emph{Kepler} or \emph{K2} C5 planet sample and the sample produced by our forward model. In overview, the Cumulative Distribution Function (CDF) of these two samples are compared and the differences are summed at each step, weighted by the location within the distribution. Theoretically, the largest differences should exist near the median of the distributions, therefore these separations are weighted less than those near the minimum and maximum values of the distribution. For a more thorough explanation of this test we refer to \cite{bab06}. The test statistic $AD\propto ln(\textrm{probability})$, measuring the probability that the two samples come from the same parent population. A similar metric is used in {\tt SysSim} \citep{he19} and a version of this metric using the Kolmogorov–Smirnov (KS; \citealt{kol33,smi48}) test is used in {\tt EPOS} \citep{muld18}. 

We compute independent measures of $AD$ for the radius and period population ($AD_R$ and $AD_P$) of our forward model distribution compared to the empirical \emph{Kepler} and \emph{K2} samples. This ensures the shape of the distributions are optimized at each step of the MCMC. However, we must also optimize the normalization of these distributions. 

Since the number of detected planets are discrete values, we rely on Poisson statistics to optimize our normalization factors. The number of planets detected by the forward model population ($N_{S}$) are compared to the empirical \emph{Kepler} or \emph{K2} C5 samples ($N_{E}$). One issue unique to the \emph{K2} data is that $N_{E}=43$ is a small discrete value, which means we do not have a good measure of the expected number of planets. In other words, the number of planets detected in the empirical sample is drawn from some Poisson distribution, but we do not know the true scale parameter ($\lambda$) of this distribution. When $N_{E}$ is large (as is the case for the \emph{Kepler} sample; $N_{E}=2,318$) it is reasonable to assume $\lambda=N_{E}$, but this assumption is less valid when $N_{E}$ is small. To account for these small number statistics we assume $N_{S}$ and $N_{E}$ come from the same Poisson distribution with an unknown $\lambda$. By multiplying these two probabilities together we get the probability of drawing $N_{E}$ and $N_{S}$ given some $\lambda$ value. Since we do not care what $\lambda$ is, just that the two drawn values came from the same distribution, we can then marginalize over the nuisance parameter $\lambda$, removing it from the equation:
\begin{equation}
\begin{split}
P(N_{S}\cap N_{E}|\lambda)= & \frac{e^{-\lambda}\lambda^{N_{S}}}{N_{S}!}*\frac{e^{-\lambda}\lambda^{N_{E}}}{N_{E}!} \\
P(N_{S}\cap N_{E})= & \int_{0}^{\infty}\frac{e^{-\lambda}\lambda^{N_{S}}}{N_{S}!}*\frac{e^{-\lambda}\lambda^{N_{E}}}{N_{E}!} d\lambda \\
P(N_{S}\cap N_{E})= & \frac{2^{-N_{S}-N_{E}-1}*\Gamma(N_{S}+N_{E}+1)}{N_{S}! * N_{E}!}
\end{split}	
\end{equation}	 
where $\Gamma$ is the gamma function. This equation ($P(N_{S}\cap N_{E})$) gives us the probability that these two discrete values come from the same Poisson distribution, without needing to know the true underlying $\lambda$. The overall difference between making the assumption that $N_{E}=\lambda$ and using the above equation is that $P(N_{S}\cap N_{E})$ allows for a slightly larger variance between values, as expected from small number statistics.

Putting together our shape metrics ($AD_R$ and $AD_P$) and our normalization probability ($P(N_{S}\cap N_{E})$) we get the log likelihood function of our posterior:
\begin{equation}
ln(\textrm{likelihood}) \propto ln(P(N_{S}\cap N_{E})) + AD_R + AD_P. 
\end{equation}
This function multiplied by the model priors provides our measure of the posterior distribution for our model parameters.

\subsection{Reliability}
As noted in \cite{bry19}, it is important that occurrence rates consider the sample reliability when optimizing their model. Without such consideration, the inferred model can be affected by our choice of planet candidacy thresholds. To account for such an effect, we calculate the reliability of each planet using the values provided by \citet{zin20} (for \emph{K2} C5) and \citet{tho18} (for \emph{Kepler}). At each step of the MCMC we draw (without replacement) each planet in our sample based on a probability corresponding to the planet's reliability. This means that the empirical \emph{K2} C5 sample will often have less than 43 planets. However, our sample has an overall reliability of 95\%, indicating that on average 41 planets will be included in the sample that is compared against the forward model sample. Comparatively, the \emph{Kepler} sample, with 2,318 planets and 98.4\% reliability, will on average be drawn with 2,281 planets.

It is important to note that the measurements of reliability provided by \citet{zin20} and \citet{tho18} are measures of systematic reliability which ignore potential astrophysical false positives. Projected double stars with small separations can dilute the transit depth, resulting in an underestimation of the planet radius \citep{cia17,ful17}. This can directly lead to an overestimation in the number of small radius planets and potentially contaminate the planet sample with eclipsing binaries. To minimize this potential contamination, \citet{tho18} looked for shifts in the centroid of the target star while the candidate was in transit. Finding such a shift provides evidence that significant flux is being contributed by a secondary source and the candidate warrants rejection from the planet sample. This metric is able to detect contaminants down to $1\arcsec$ separations \citep{bry13}. Unfortunately, such measurements are more difficult to establish for \emph{K2}, where spacecraft systematics are constantly shifting the centroid. Consequently, \citet{zin20} relied on the \emph{Gaia} DR2 data to minimize these false positives, which provides contaminant detections down to $1\arcsec$ separations \citep{Zie18}. While unique in methodology, both catalogs are robust to contaminants wider than $1\arcsec$ separations. However, contaminants within a $1\arcsec$ separation (largely gravitationally bound binaries; \citealt{Horch2014}) will remain undetected and reduce the overall reliability of these catalogs. While such corrections are essential for accurate occurrence measurements, our sample is limited to planets with $R\le6.7R_{\Earth}$, minimizing contamination from eclipsing binaries \citep{fre13}. Work by \citet{mat18} found that \emph{K2} planet hosts have a binarity rate of $23\pm5\%$ and \citet{fur17} found a similar value of 30\% for \emph{Kepler} planet hosts, but implementing such information into an occurrence rate is beyond the scope of this paper and therefore ignored.

\section{Results}
\label{sec:Result}

In this section we compare three different models against the empirical \emph{K2} C5 planet sample: the \emph{K2} model, which only uses the 43 planets in our sample to fit the seven population parameters; the \emph{Kepler} model, which uses the 2,318 \emph{Kepler} planet candidates to fit the seven population parameters; and the \emph{K2} w/ \emph{Kepler} model, which uses the six shape parameter posterior distributions derived by the \emph{Kepler} model and fits the normalization factor ($f$) using the 43 \emph{K2} C5 planet candidates.

\begin{deluxetable*}{lccccccc}
\tablecaption{The resulting best fit parameters of our forward model optimization. The \emph{K2} w/ \emph{Kepler} model uses the same shape posteriors as that of the \emph{Kepler} model. In Figure \ref{fig:posterior} we plot the posterior distributions for the $f$ values provided here.
\label{tab:param}}
\tablehead{\colhead{Model} & \colhead{$\alpha_1$} & \colhead{$R_{br}$} & \colhead{$\alpha_2$} & \colhead{$\beta_1$} & \colhead{$P_{br}$} & \colhead{$\beta_1$} & \colhead{$f$} \\ 
\colhead{} & \colhead{} & \colhead{($R_{\oplus}$)} & \colhead{} & \colhead{} & \colhead{(days)} & \colhead{} & \colhead{(planets/star)} } 

\startdata
\textbf{\emph{Kepler}} & $-1.61^{+0.17}_{-0.14}$ & $3.03^{+0.38}_{-0.37}$ & $-6.56^{+1.77}_{-5.58}$ & $0.91^{+0.19}_{-0.17}$ & $6.83^{+1.59}_{-1.26}$ & $-0.59^{+0.15}_{-0.18}$ & $1.10^{+0.05}_{-0.05}$ \\
\textbf{\emph{K2} C5} & $-0.38^{+1.78}_{-1.29}$ & $2.98^{+1.22}_{-1.29}$ & $-6.99^{+3.28}_{-10.01}$ & $1.66^{+6.49}_{-1.26}$ & $6.91^{+2.07}_{-4.03}$ & $0.15^{+0.71}_{-0.90}$ & $1.00^{+1.07}_{-0.51}$ \\
\textbf{\emph{K2} w/ \emph{Kepler}} & \nodata & \nodata & \nodata & \nodata & \nodata & \nodata & $0.77^{+0.21}_{-0.20}$ \\
\enddata

\end{deluxetable*}

\subsection{\emph{K2} C5 Model}
Using the 43 planets in our C5 FGK sample, we optimize the seven population parameters. Since the number of planets is only six times greater than the number of parameters, the uncertainties in our estimates for this model are very large. Nevertheless, we can still provide some measure of the population parameters.

\begin{figure*}[!ht]
\centering \hfill \includegraphics[height=6.4cm]{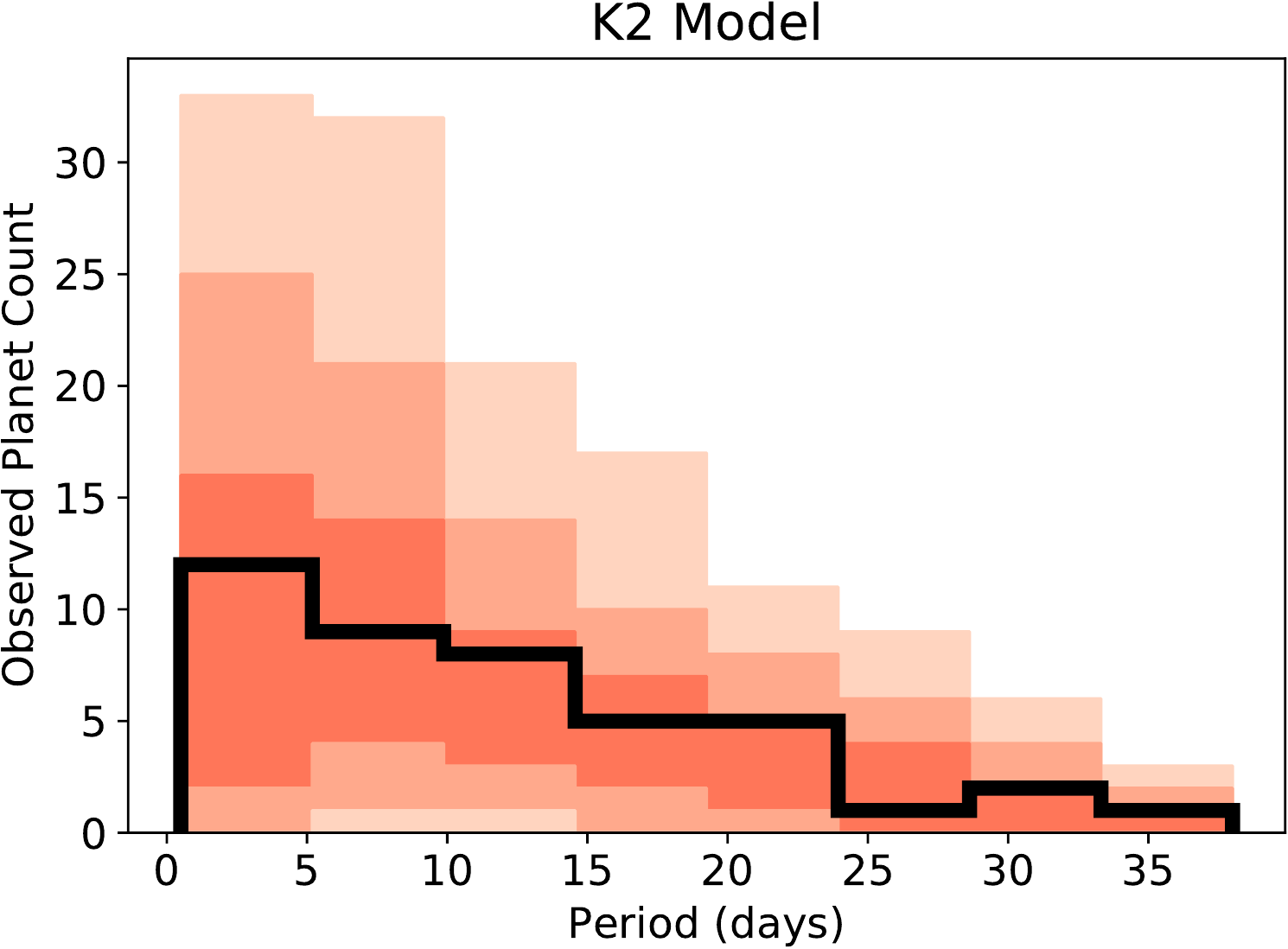}
\centering \hfill \includegraphics[height=6.4cm]{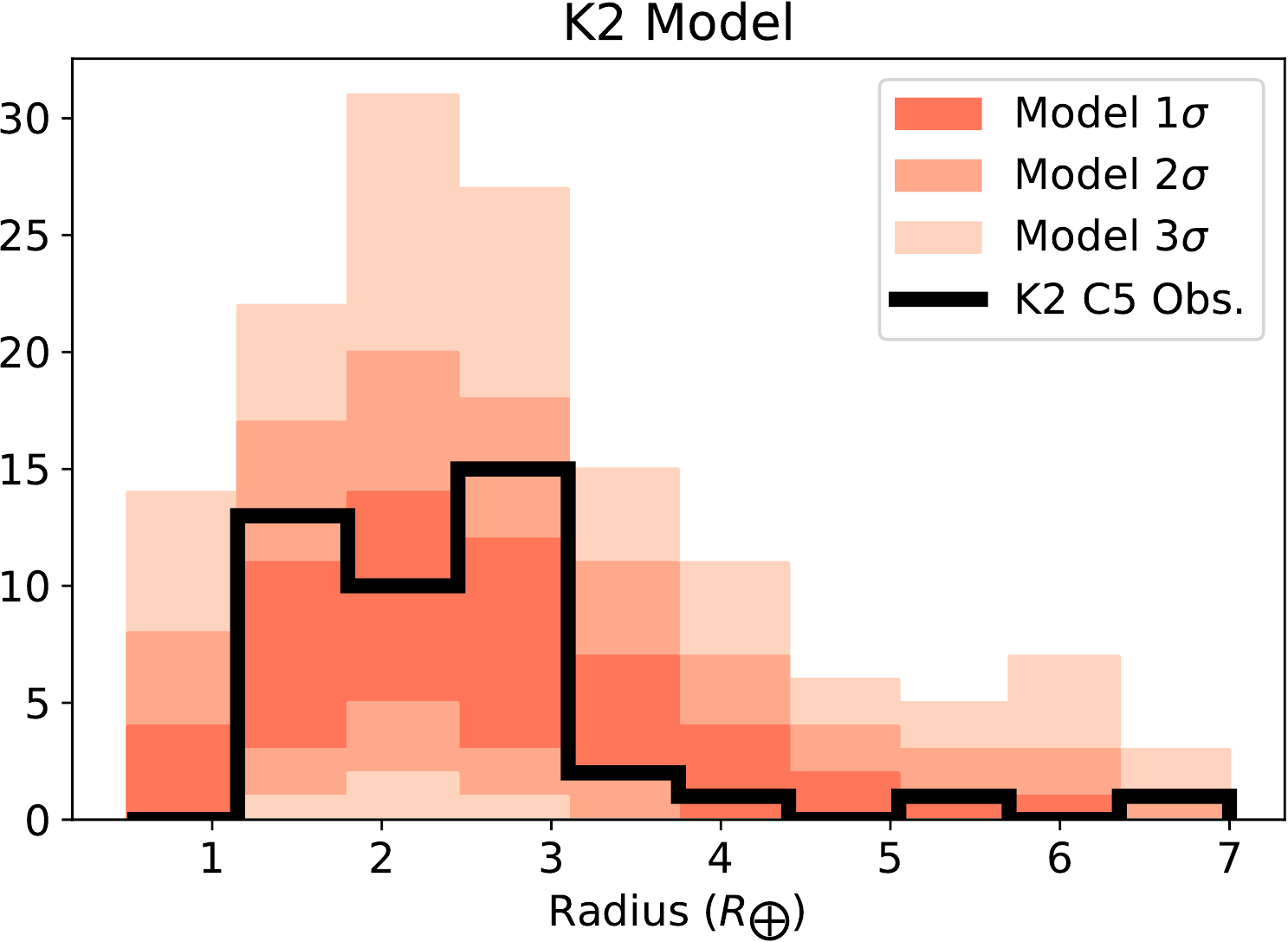}
\centering \hfill \includegraphics[height=6.4cm]{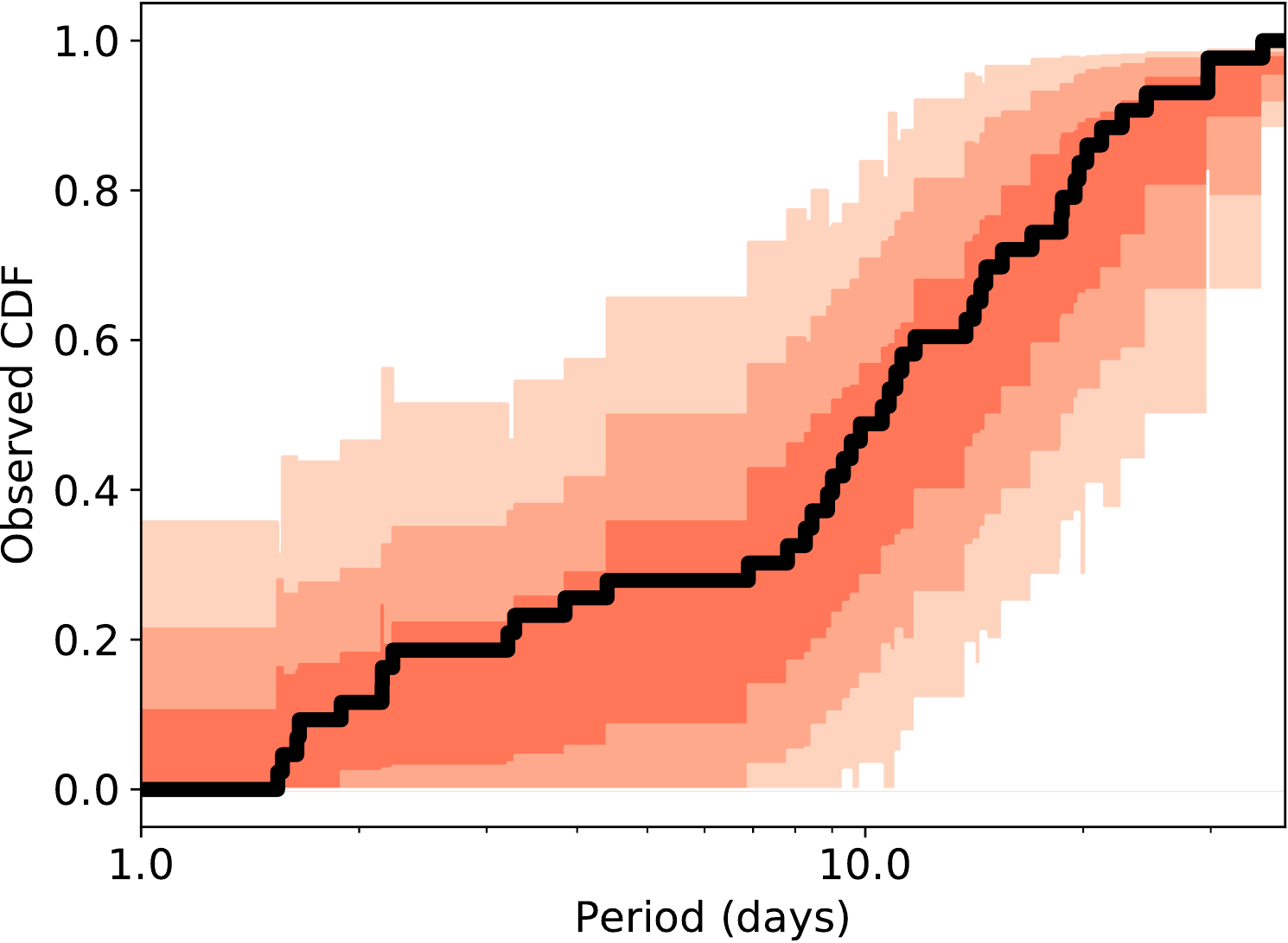}
\centering \hfill \includegraphics[height=6.4cm]{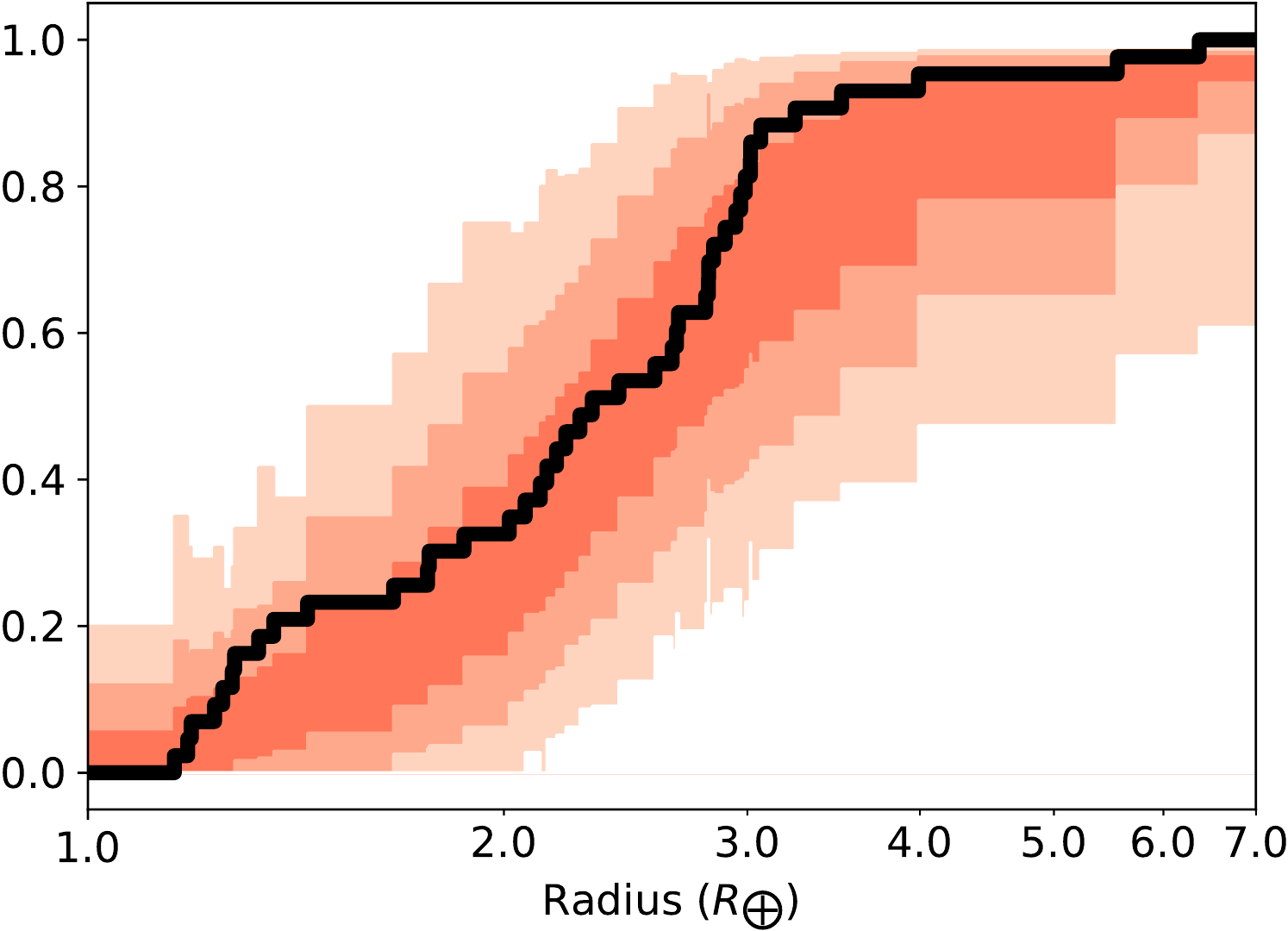}
\caption{\textbf{Top} the observed radius/period distributions for the \emph{K2} C5 data set versus the expected detections from the best fit broken power-law model, binned to show that the normalization factor correctly matched the observed number of planets. \textbf{Bottom} the observed CDF for the \emph{K2} C5 data plotted against the expected detections from the best fit model, indicating a good match of the distribution shapes. The colored regions reflect the 68, 95, and 99.7 percentiles found after sampling with the model parameters 10,000 times. \label{fig:k2only}}
\end{figure*}

In Table \ref{tab:param} we provide the results of our MCMC for the population parameters using only the \emph{K2} C5 data. To ensure the model reflects the data, we look at the model distributions in Figure \ref{fig:k2only}. It is clear that the model subjected to the selection effects of our pipeline follows the shape (CDF) of the period and radius distributions. Overall, these model parameters find the same trend that previous occurrence rates have noted: a constantly decreasing radius distribution ($\alpha_1$ and $\alpha_2 < 0$; \citealt{bur15,muld18}) and a peaked period distribution around 10 days ($P_{\textrm{br}}\sim 10$ days; \citealt{you11,how12,muld19}). We can also see in Figure \ref{fig:k2only} that this model produces a proper normalization of the data, producing an expectation of $34^{+11}_{-11}$ detected planets. This normalization is controlled by the overall occurrence parameters ($f$), which indicates our stellar sample should host on average $1.00^{+1.07}_{-0.51}$ planets per star. It is not surprising that we find a value near one, as the 38 day period limit of this study removes a significant fraction of system multiplicity. However, it is likely that this number is a lower estimate as we have assumed no mutual inclination between planets for simplicity. The real strength of this value is that it allows us to make a comparative statement to that of the \emph{Kepler} model. 

\subsection{\emph{Kepler} Model}

Using our \emph{Kepler} planet and stellar sample we report the results of our \emph{Kepler} model optimization in Table \ref{tab:param}. Clearly the larger planet and stellar samples help reduce the overall uncertainty in our population parameters. Unsurprisingly, the \emph{Kepler} model shape parameters are in agreement with previous \emph{Kepler}-centric studies (i.e. \citealt{zin19}; $\alpha_1=-1.65$, $R_{\textrm{br}}=2.66$, $\alpha_2=-4.35$, $\beta_1=0.76$, $P_{\textrm{br}}=7.09$, and $\beta_2=-0.64$), which uses a larger planet sample that spans the full 500 day period range of the \emph{Kepler} data set. By cutting the data and simplifying the processing to match that of \emph{K2} C5, we can now look for potential differences and similarities between these two independent samples. 

\begin{figure*}[!ht]
\centering \hfill \includegraphics[height=6.4cm]{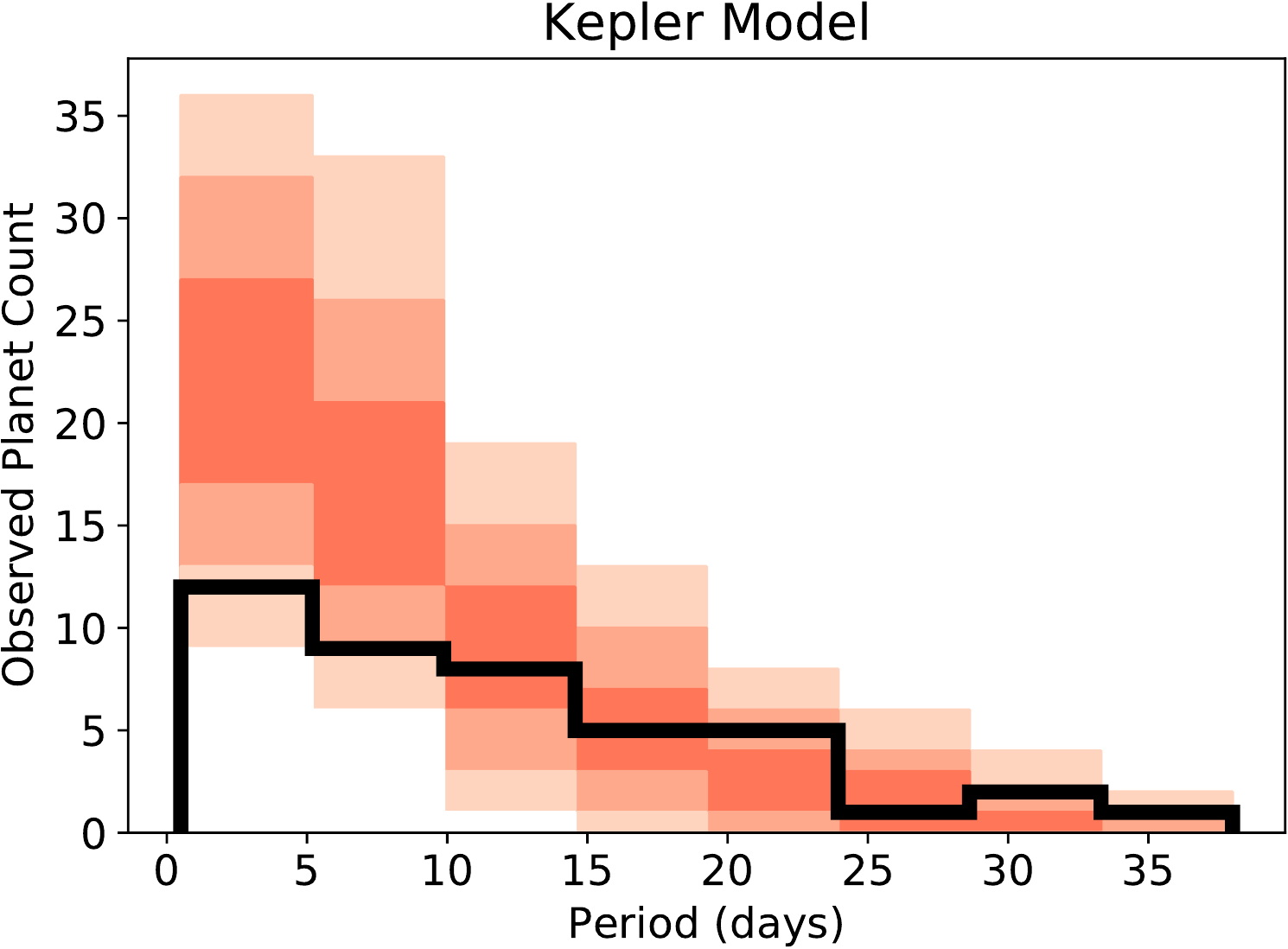}
\centering \hfill \includegraphics[height=6.4cm]{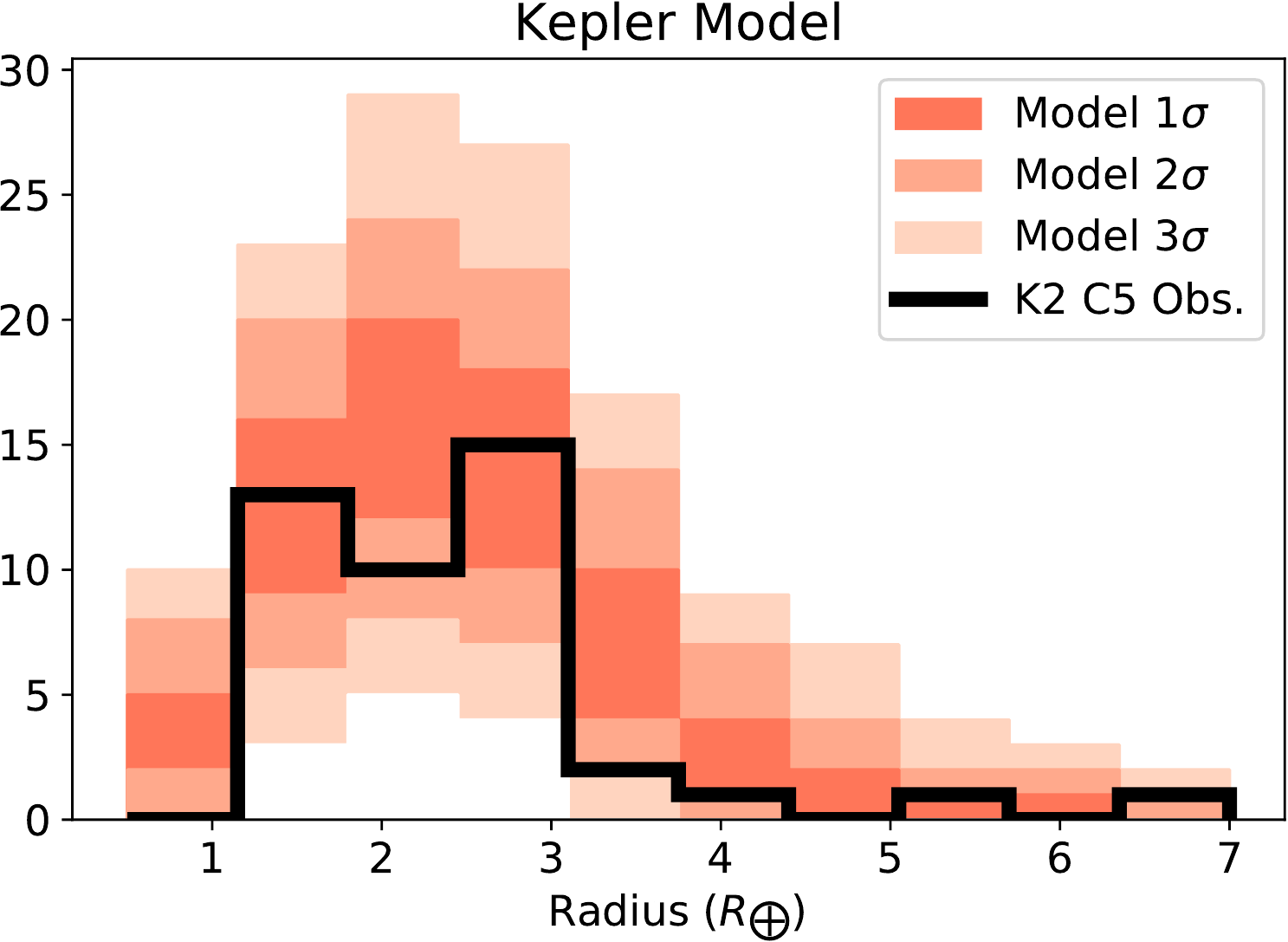}
\centering \hfill \includegraphics[height=6.4cm]{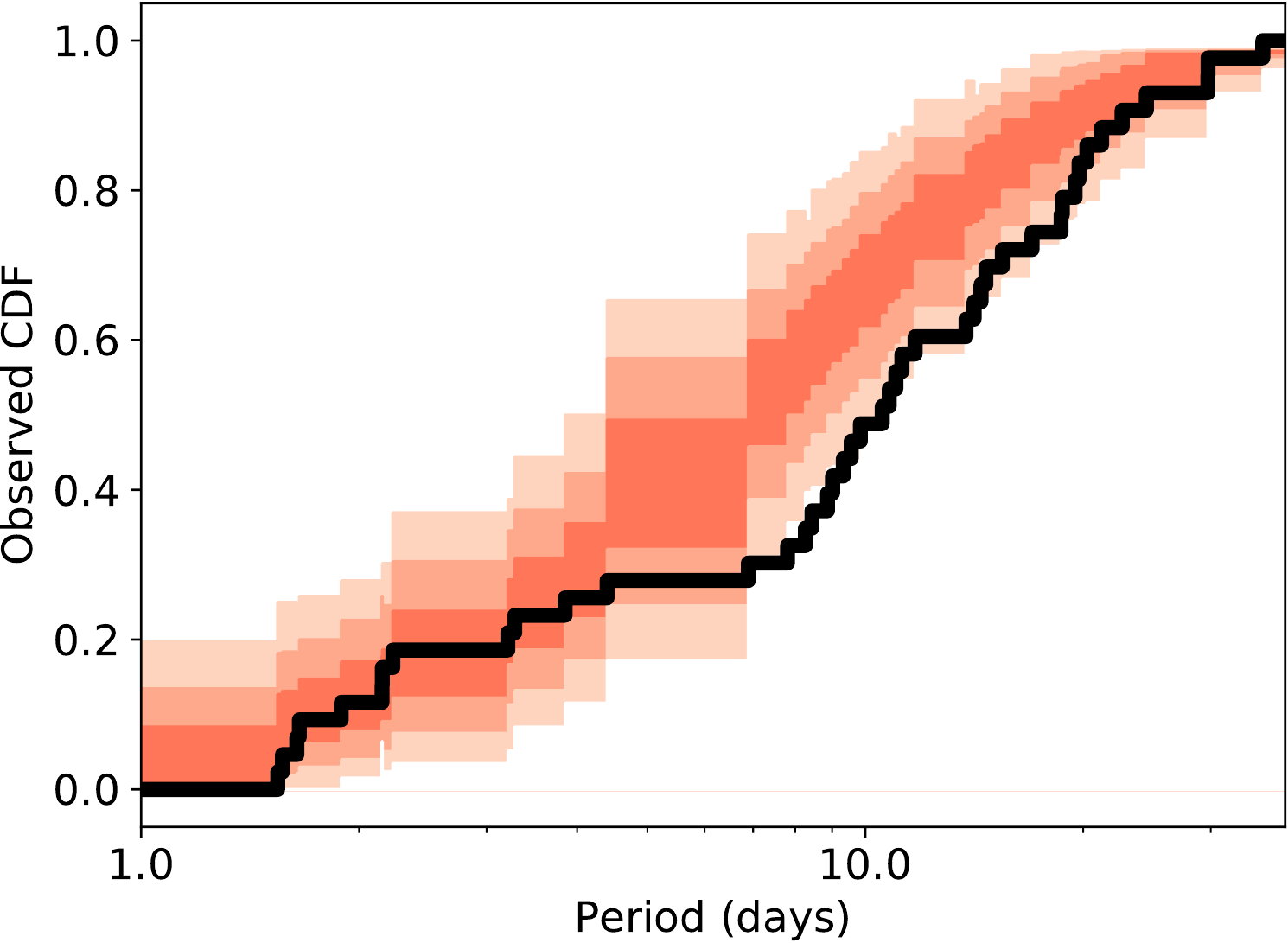}
\centering \hfill \includegraphics[height=6.4cm]{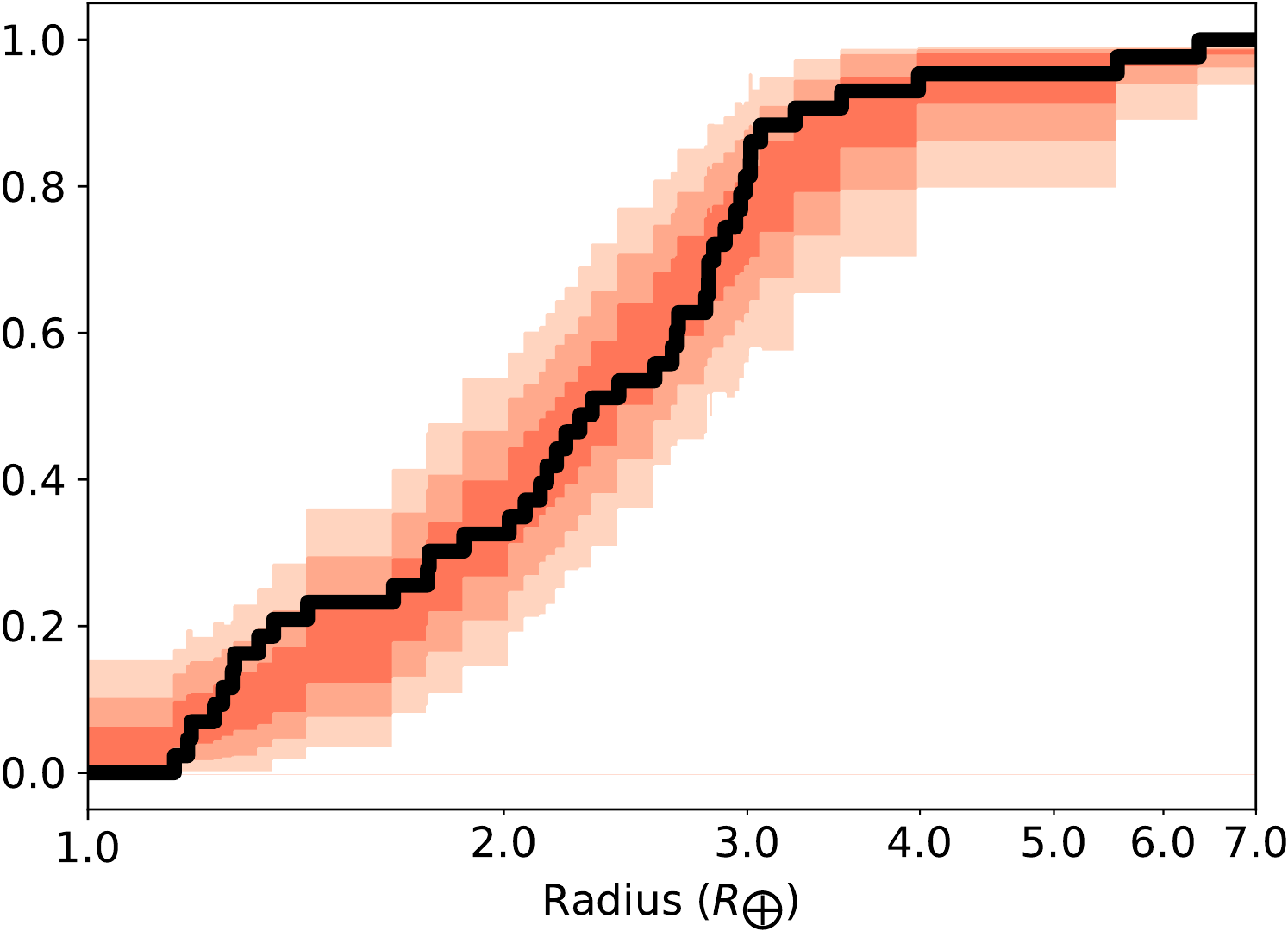}
\caption{\textbf{Top} the observed radius/period distributions for the \emph{K2} C5 data set versus the expected detections from the \emph{Kepler} broken power-law model. \textbf{Bottom} the observed CDF for the \emph{K2} C5 data plotted against the expected detections from the best fit \emph{Kepler} model, indicating a reasonable match of the distribution shapes. The colored regions reflect the 68, 95, and 99.7 percentiles found after sampling with the model parameters 10,000 times. \label{fig:kepleronly}}
\end{figure*}

In Figure \ref{fig:kepleronly} we present the detected sample for the \emph{Kepler} model population parameters given the selection effect of the \emph{K2} C5 pipeline. This allows us to see how the \emph{Kepler} planet population would have been detected by \emph{K2}. It is apparent that the \emph{K2} C5 sample is within $3\sigma$ of nearly all aspects of the \emph{Kepler} population model. Furthermore, in Table \ref{tab:param} all of the \emph{K2} C5 inferred population parameters are well within $1\sigma$ of their corresponding \emph{Kepler} parameters. This indicates that the \emph{K2} C5 sample is not statistically different from the \emph{Kepler} population. We caution that a lack of statistical significance does not ensure true similarity; a much larger planet sample will be needed to make such a claim. However, it appears that any differences that do exist will be minor. If we consider the overall expected number of detected planets, this model predicts the detection of $58\pm8$ planets (compared to the 43 in our \emph{K2} C5 sample). This similarity further indicates a lack of uniqueness among these samples.

Although differences in the populations are statistically insignificant, the uncertainties in the \emph{K2} C5 data remains high. Should subtle differences exist, a larger \emph{K2} sample would be needed for detection. One method of improving the fit for the overall occurrence factor ($f$) would be to reduce the number of model parameters. It is likely that the same formation mechanisms are at play in both the \emph{Kepler} and \emph{K2} samples, thus the overall shape of these distributions should be similar and information from \emph{Kepler} can be used to help fit the \emph{K2} data.

\subsection{\emph{K2} w/ \emph{Kepler} Model}
\label{sec:K2wKepler}
To minimize the number of parameters in the \emph{K2} C5 model we use the posterior values from the \emph{Kepler} shape distribution for optimization. At each step of the MCMC a set of shape values are drawn from the posterior samples inferred by the \emph{Kepler} model. Drawing values in a set ensures we maintain any dependence between parameters. The only parameter that is allowed to roam is the occurrence factor ($f$), which normalizes the distribution. By reducing the parameter space search we can produce a sharper estimate of the overall planet occurrence in \emph{K2} C5.

\begin{figure}
\centering \includegraphics[width=8.5cm]{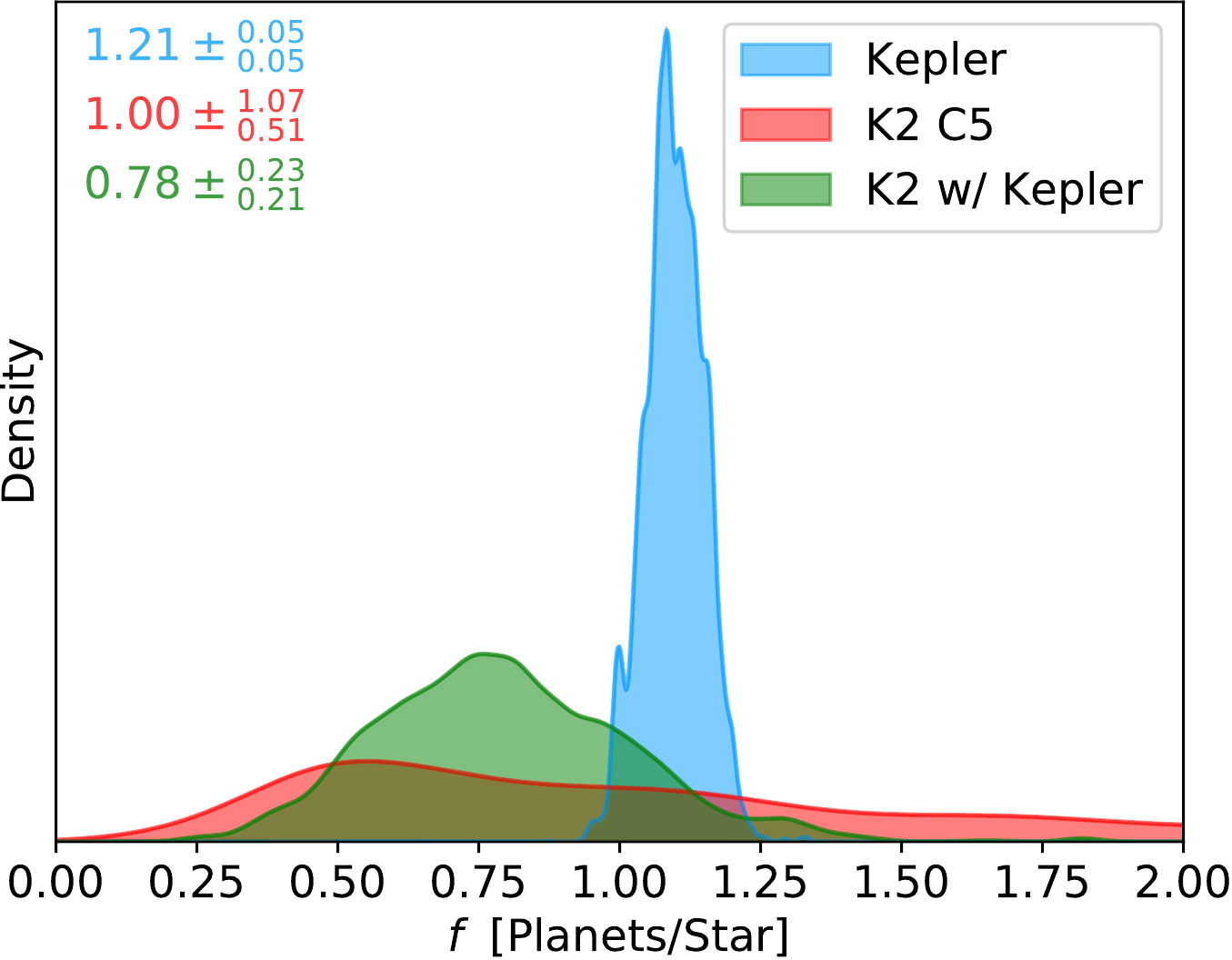}

\caption{The posterior distributions for the occurrence factors ($f$) derived from our planet samples. The 50th, 16th and 84th percentiles have been listed in the upper corner of the plot for each value. Clearly, there is significant overlap between all of these values. 
\label{fig:posterior}}
\end{figure}

We find that reducing the parameter search produces a 22\% lower occurrence factor ($f=0.78^{+0.23}_{-0.21}$) and a 70\% reduction of the parameter uncertainty. Although reduced even further, we still do not find a statistically significant difference from that of the \emph{Kepler} model values ($f=1.21\pm0.05$). In Figure \ref{fig:posterior} we show the posterior distributions for our three model occurrence factors and their significant overlap, providing further evidence that differences in these two independent populations will be subtle, if existent. 

Since both \emph{K2} models produce statistically indistinguishable occurrence factors (when compared to the \emph{Kepler} model), we can use the posteriors to bound the differences that could exist and remain undetected by our study. By considering the $3\sigma$ posterior values for $f$, we can bound $\Delta f=f_\textrm{Kepler}-f_\textrm{K2}$ to a range of $[-4.47,0.99]$ planets per star for the \emph{K2} C5 model and $[-0.86,0.98]$ planets per star for the \emph{K2} w/ \emph{Kepler} model. Clearly, these bounds remain rather weak due to the large uncertainty in \emph{K2} occurrence, but both models are able to rule out $\Delta f=1$. Should the true $f_\textrm{Kepler}$ value be greater than the $f_\textrm{K2}$ value--as expected by the median posterior values--the difference will be less than one planet per star in the parameter space of this study.

\subsection{\emph{K2} and \emph{Kepler}}
\label{sec:K2/Kepler}
Since we do not find an overall statistical difference between these two populations, it seems reasonable to combine these two samples to improve the overall model fit. However, appropriately combining the \emph{Kepler} DR25 sample with the \emph{K2} sample is difficult because both empirical samples have unique selection effects that need to be taken into account. Fortunately, Bayesian analysis provides a direct way of implementing this type of knowledge. Since the \emph{Kepler} sample is an independent measurement of the population parameters, we can use the inferred \emph{Kepler} posterior values as priors for our model. This will essentially update the posteriors given the new \emph{K2} data, pulling the posteriors closer to the true population parameters. In a preliminary study, this Bayesian analysis was carried out for the \emph{K2} C5 sample, but the C5 sample only added an additional 43 planets to the \emph{Kepler} sample of 2,318 planets (a 1.9\% increase), providing little influence on the overall population parameters. Once a larger portion of the \emph{K2} Campaigns have been processed we will use this methodology to combine data across missions.

\section{Deviations from the Model}
\label{sec:radgap}
In Figures \ref{fig:k2only} and \ref{fig:kepleronly} it is clear that the broken power-law model for the planet radii fails to replicate the gap seen in the radius sample near $R=2R_{\Earth}$. Currently, the cause of this gap remains unclear. \citet{lop13} and \citet{owe17} suggest that the gap is caused photoevaporation, while \cite{gup18} indicates such a gap could be caused by core-powered mass loss. Regardless, this gap has been noted in both the \emph{Kepler} \citep{ful17} and \emph{K2} \citep{har20} planet samples, indicating an underlying formation mechanism is at play. As we continue to increase the known planet sample, we will be able to better constrain the underlying mechanism \citep{loy20}.

Without a well defined model available for this gap, it remains difficult to recreate in our population analysis. We acknowledge that our model does not address this issue. We also attribute the increased planet occurrence near $2.75R_{\Earth}$ and the subsequent decrease in planet occurrence seen near $3.1R_{\Earth}$, which appears to deviate from a power-law model in the empirical \emph{K2} C5, to this lack of a well defined radius gap model.

Both photoevaporation and core-powered mass loss predict a period dependency to this gap. In the \emph{Kepler} data set, which included planets with periods of 0.5--500 days, the effect of the radius gap is almost completely washed out when considering the period distribution on its own. However, our \emph{K2} C5 sample is limited to periods of 0.5--38 days, where most of the detected planets have periods less than 10 days. Additionally, almost all radius gap models intersect at the junction of $R=1.7R_{\Earth}$ and $P=7$ days, where the gap is most prominent \citep{mac19}. Combining these two facts, our smaller under-sampled slice of the observable exoplanet period population is prone to gaps in the period distribution. We see such a gap in our empirical sample between 4--7 days, where the period distribution appears to deviate from the power-law (see Figure \ref{fig:kepleronly} Bottom). We therefore conclude that this apparent gap is not meaningful.

An additional complexity was introduced by \cite{mil17} and \cite{wei18}, who showed that planets are not independent within a given system. The ``Peas in a Pod'' result shows that planets within a given system tend to have similar sizes. Modeling this type of radius correlation remains difficult for population analysis. \cite{he19} proposed a clustered point process model for dealing with system radius similarity. However, a proper accounting for such similarities would require a thorough understanding of the multiplicity completeness, which is currently not available for the \emph{K2} data set. Furthermore, accounting for these empirical features requires additional parameters, which we have tried to minimize due to the small size of the \emph{K2} C5 sample. Fortunately, only seven planets in our \emph{K2} sample exist in multi-planet systems, therefore the overall effect of this correlation on the inferred population is small. Of the \emph{Kepler} sample, 1,020 planets are part of multi-planet systems, and thus accounting for such correlations becomes important. We have ignored such issues here in order to minimize the number of parameters and to make the analysis between \emph{Kepler} and \emph{K2} data equivalent.

\section{Effects of Metallicity}
\label{sec:metal}
In Section \ref{sec:sample} we discussed the differences between the \emph{Kepler} and \emph{K2} C5 stellar samples. The most notable difference is the metallicity of these two samples. The \emph{Kepler} sample represents a more metal-rich sample than the \emph{K2} C5 stellar sample. When considering the trends observed in \citet{pet18}, we should expect to find more sub-Neptunes (1.7--4$R_{\Earth}$) in the metal-rich \emph{Kepler} sample. 

While we do not directly consider this metallicity effect, we can discuss the expected consequences in our results. Since a larger number of sub-Neptunes should be found in the \emph{Kepler} sample, we expect this to increase the overall occurrence of planets in the \emph{Kepler} model. Although statistically insignificant, we do find a slight increase in the occurrence of planets produced by our \emph{Kepler} model ($\Delta f=0.21\pm1.07$ planets per star) compared to the \emph{K2} C5 model. This difference is amplified even further when considering the \emph{Kepler} and \emph{K2} w/ \emph{Kepler} model ($\Delta f=0.33\pm0.22$ planets per star), producing a $1.5\sigma$ difference.

Additionally, we would also expect the $\alpha_2$ population parameter to be slightly inflated for the \emph{Kepler} sample as there would be a greater number of sub-Neptunes in the radius range this parameter spans. Again, we find a statistically insignificant increase between our models ($\Delta \alpha_2=0.4\pm6.47$). While the differences observed are not able to confirm the findings of \citet{pet18}, they are in agreement with the expectations of such a metallicity effect. We leave a more thorough consideration of this effect for future studies when a larger \emph{K2} sample is available.  

\section{Conclusions}
\label{sec:conclusion}

We used the \emph{K2} C5 fully automated detection pipeline data set to determine the underlying population of planets around FGK dwarfs in the C5 field. In doing so we were able to infer an overall occurrence of $1.00^{+1.07}_{+0.51}$ planets per star in the parameter space of this study. When we compared the population parameters to those of the best-fit \emph{Kepler} model, we found that all values are well within a $\sim$1$\sigma$ difference, including the overall occurrence factor ($1.10\pm0.05$). Even when using the \emph{Kepler} model shape parameters to improve the optimization of the \emph{K2} C5 occurrence factor, we found only a $1.5\sigma$ decrease in planet occurrence in the \emph{K2} C5 field. Despite the C5 field probing a different region of the Galaxy, we infer a population that appears consistent with the \emph{Kepler} sample. This indicates that our knowledge of the \emph{Kepler} field could potentially be extrapolated to a larger part of the Galaxy. 

Using Bayesian priors, we also discussed a methodology for combined \emph{K2} \& \emph{Kepler} mission data to carry out a Galactic transiting exoplanet occurrence rate. With the currently available data, this analysis would be heavily biased toward the \emph{Kepler} field data (2,318 \emph{Kepler} planets versus 43 planets from C5). A more rigorous Galactic survey would sample various regions of the local Galaxy, with a similar data span at each of them. Fortunately, the \emph{K2} mission did just that, and as more campaigns are fully processed with the automated pipeline we will use this methodology to calculate a representative Galactic occurrence rate for planets.

Finally, we showed that the \emph{K2} stellar sample is metal-poor compared to the \emph{Kepler} stellar sample, but we were unable to find statistical differences in our models. Findings of model similarity may reduce the role metallicity plays in planet formation, however our results found a weak increase in the occurrence of planets in the \emph{Kepler} field. This trend seems to indicate that a larger planet sample--or a more substantial sample metallicity difference--is needed to confirm/refute the importance of metallicity in planet formation. Once the entire \emph{K2} planet sample is made available, a more thorough consideration of metallicity effects can be achieved.

\section{Acknowledgements}
The simulations described here were performed on the UCLA Hoffman2 shared computing cluster and using the resources provided by the Bhaumik Institute. This research has made use of the NASA Exoplanet Archive and the Exoplanet Follow-up Observation Program website, which are operated by the California Institute of Technology, under contract with the National Aeronautics and Space Administration under the Exoplanet Exploration Program. This paper includes data collected by the \emph{Kepler} mission and obtained from the MAST data archive at the Space Telescope Science Institute (STScI). Funding for the \emph{Kepler} mission is provided by the NASA Science Mission Directorate. STScI is operated by the Association of Universities for Research in Astronomy, Inc., under NASA contract NAS 5–26555.

\facilities{Exoplanet Archive, Kepler, K2}

\bibliography{./bib.bib}\setlength{\itemsep}{-2mm}



\end{document}